\documentclass[]{pasj00}
\usepackage{lscape}
\RelativeFigurePath{./}
\begin{document}

\pagestyle{sample}
\SetRunningHead{T. Sato et al.}{Suzaku Observations of the Coma Cluster}

\title{Suzaku Observations of Iron K-lines from
the Intracluster Medium of the Coma Cluster}

\author{Takuya \textsc{Sato}\altaffilmark{1}, 
Kyoko \textsc{Matsushita}\altaffilmark{1}, 
Naomi \textsc{Ota}\altaffilmark{2}, 
Kosuke \textsc{Sato}\altaffilmark{1}, 
Kazuhiro \textsc{Nakazawa}\altaffilmark{3}, 
and Craig L. \textsc{Sarazin}\altaffilmark{4}}

\altaffiltext{1}{
Department of Physics, Tokyo University of Science, 
1-3 Kagurazaka, Shinjuku-ku, Tokyo 162-8601 }
\email{j1209703@ed.kagu.tus.ac.jp; matusita@rs.kagu.tus.ac.jp}
\altaffiltext{2}{
Department of Physics, Nara Women's University, 
Kitauoyanishi-machi, Nara, Nara 630-8506 }
\altaffiltext{3}{
Department of Physics, University of Tokyo, 
7-3-1 Hongo, Bunkyo-ku, Tokyo 113-0033 }
\altaffiltext{4}{
Department of Astronomy, University of Virginia, 
P.O. Box 400325, Charlottesville, VA 22904-4325 }

\Received{2011/06/09}
\Accepted{2011/08/29}
\Published{---}

\maketitle

\begin{abstract}
The Coma cluster was observed with an X-ray Imaging Spectrometer 
(XIS) onboard Suzaku in six pointings, including the central 
X-ray peak region, the 14$'$ west offset region, 30$'$ and 34$'$ north-west 
offset regions, and 44$'$ and 60$'$ south-west offset regions.
Owing to its lower background level, Suzaku has better sensitivity 
to Fe K$\alpha$ lines than other satellites. 
Using precise Fe line measurements, we studied 
the temperature structure, possible bulk motions, and iron 
abundance distributions in the intracluster medium (ICM). The observed spectra were 
well-represented by a single-temperature model, and two- or three- 
temperature model did not improve $\chi^2$ substantially. 
The temperature, derived from K$\alpha$ line ratios of H-like 
and He-like Fe, agree with those derived from the single-temperature model. 
Because the line ratio is a steep function of temperature,
the consistency supports the accuracy of temperature measurements conducted with Suzaku.
Within the 34$'$ region, the redshift derived from the central energy of 
the He-like Fe line is consistent with that from optical 
observations, within a calibration error of 18 eV or 
818~km~s$^{-1}$ in the line of sight. 
This value is smaller than the sound velocity of ICM, which is 1500~km~s$^{-1}$.
The central energy of Fe lines at
the 44$'$ offset region around the NGC~4839 subcluster is also consistent with those within the 34$'$ region. 
These results on the temperature and velocity structure 
suggest that the core of the cluster is in a relaxed state,
and non-thermal electrons relevant to the radio halo 
are accelerated by intracluster turbulence rather than large-scale shocks. 
Fe abundance is almost constant at 0.4 solar within the 34$'$ region, and decreases 
with radius. This value is slightly lower than 
those of other clusters, which means the gas have been 
mixed well during a past merger associated with the growth of the cluster.
\end{abstract}

\section{Introduction}

Clusters of galaxies are thought to grow into larger systems through 
complex interactions between smaller systems. Signatures of
merging events include temperature and density 
inhomogeneities and bulk motions in the intracluster medium (ICM). 
Numerical simulations predict simplified bulk 
motions with a substantial fraction of virial velocity 
($\sim$1000~km~s$^{-1}$), lasting several Gyr after each merging 
event (e.g., \cite{roettiger1996}; \cite{norman1999}). Measurements 
of the ICM temperature and velocity structure, therefore, provide 
crucial information for understanding the evolution of clusters.

Velocity measurements of bulk motion and turbulence are essential 
for mass estimation and cosmological studies. If ICM has a 
significant bulk velocity compared to its thermal velocity, the 
associated non-thermal pressure would threaten the assumption of 
hydrostatic equilibrium in deriving the total gravitational mass in a 
cluster. For example, factors of 2--3 discrepancies between the X-ray and 
lensing mass estimates in some objects (e.g., \cite{ota2004}; 
\cite{hattori1999}) could be partly due to this effect.
\citet{Tamura2011} discovered a significant bulk motion in Abell~2256, 
which is a well known merging cluster. \citet{schuecker04} performed a 
Fourier analysis of XMM-Newton data of the Coma cluster, which 
revealed the presence of a scale-invariant pressure fluctuation 
ranging between 29 and 64 kpc, and $\sim$10\% of the total pressure 
in turbulent form. However, in relaxed clusters, significant bulk motions and 
turbulence have not been detected. 
On the basis of ASCA and Chandra data,
\citep{dupke2001,dupke2006} claimed a large velocity gradient of 
$\sim$2400 km~s$^{-1}$  in the Centaurus cluster, which is a relaxed 
cluster with a cool core. However, \citet{ota07} found a negative 
result of $|\Delta v|<1400$~km~s$^{-1}$ on the basis of Suzaku observations 
of the Centaurus cluster. \citet{ksato08} also searched for 
possible bulk motions in the AWM~7 cluster with Suzaku, and observed no 
significant bulk motions. The upper limit of gas velocity was 
$|\Delta v|<$ 2000~km~s$^{-1}$. 
\citet{fujita2008} do not detect the variation of the redshift of the ICM 
in the Ophiuchus cluster, and the upper limit of the velocity
difference is 3000~km~s$^{-1}$.
\citet{sugawara2009} studied the ICM 
flow in Abell~2319 with Suzaku, and detected no velocity differences within 
the observed region. Using spectra with a Reflection 
Grating Spectrometer (RGS) onboard XMM-Newton, \citet{Sanders2011} 
provided upper limits on turbulence velocities
in cluster core regions: at least 15 sources 
had less than 20\% of the thermal energy density in turbulence.

The Coma cluster ($z=0.0231$), also known as Abell~1656, is one of 
the most studied clusters \citep{biviano1998} and has been observed 
at all wavelengths from radio to hard X-ray bands. However, the 
physical state, particularly the dynamical state, of the Coma cluster 
has not been completely understood. Measurements of the velocity 
dispersion and distribution of galaxies belonging to the Coma 
cluster constrain the dynamical history of the Coma cluster. 
\citet{fitchett1987} found that two central galaxies, NGC~4889 
and NGC~4874 have a significant difference in velocity, which is 
evidence of a recent merger. \citet{colless1996} argued that NGC~4874 
was the original dominant galaxy of the main cluster, and that
NGC~4889 belonged to a subgroup, that recently merged with the main 
cluster. A substructure associated with the galaxy, NGC 4839, located 
40$'$ south-west of the cluster core, was found by 
\citet{mellier1988} and \citet{merritt1994}. The existence of this 
subcluster was confirmed by \citet{colless1996}.

The thermal X-ray emission of the Coma cluster has been observed 
with several X-ray satellites. \citet{briel1992} constructed an 
X-ray map of the Coma cluster using ROSAT All-Sky-Survey data and 
determined the X-ray surface brightness profile of the cluster 
out to a radius of roughly 100$'$. 
The temperature maps of clusters provide knowledge about
the history of past subcluster mergers. On the basis of
ASCA observations, \citet{honda1996} and \citet{watanabe1999} 
found that ICM is not isothermal. 
\citet{arnaud01} used XMM-Newton to study the temperature structure in the central 
region of the Coma cluster. The projected temperature distribution 
around NGC~4889 and NGC~4874 is remarkably homogeneous, which suggests 
that the core is mostly in a relaxed state.
 Except at the center,
the temperature decreases slightly with radius. 
A cool filament of an X-ray emission 
in the direction of the galaxy NGC~4911, which is located in the south-east 
of the cluster center, was detected with Chandra \citep{vikhlinin1997}
and XMM \citep{arnaud01}.
ROSAT observed a substructure around the NGC~4839 subgroup \citep{white1993},
and
a hot region in the direction of NGC~4839 was observed by \citet{arnaud01}.
By XMM-Newton observations, \citet{neumann2001} 
found compelling evidence for the subgroup around NGC~4839
 was on its first infall into the Coma cluster, 
but it had not passed its core yet. This 
interpretation is different from that of \citet{burns1994}, who 
suggested that the NGC~4839 group was moving out of the cluster after 
having already passed through the Coma cluster.

Non-thermal electrons, observed via diffused radio synchrotron emission, 
have been detected in more than 50 clusters, and all of them are undergoing 
mergers (e.g., \cite{buote2001}; \cite{schuecker2001}). 
The Coma cluster also has a cluster-wide synchrotron radio halo, emitted 
by relativistic electrons due to merging \citep{feretti1998}.
Recent non-thermal detections have been claimed by \citet{rephaeli2002} 
with RXTE and by \citet{fusco2004,fusco1999} with BeppoSAX, 
although the latter detection is controversial 
\citep{rosetti2004,fusco2007}.
 Long 
observations ($\sim$1 Ms) by INTEGRAL have imaged extended diffuse 
hard X-ray emission from the Coma cluster, although it was found to be 
completely consistent with thermal emission
\citep{renaud2006, eckert2007, lutovinov2008}.
Using the temperature map of ICM obtained with  the PN detector
onboard XMM-Newton, \citet{wik09} analyzed data obtained from the Suzaku Hard X-ray 
Detector (HXD) \citep{takahashi07} and derived the strongest upper limit
of non-thermal emission.

This paper reports results from six Suzaku observations of the Coma 
cluster out to 60$'$, $\simeq$1.5 Mpc, conducted using the X-ray Imaging 
Spectrometer (XIS) \citep{koyama07} onboard Suzaku \citep{mitsuda07}.
Owing to its lower background level, Suzaku has better sensitivity 
to the iron K$\alpha$ lines than previous satellites.
In addition, the accurate calibration 
of XIS allows us to measure the velocity of ICM precisely. 

We used the Hubble constant, $H_{\rm 0} = 70$~km~s$^{-1}$~Mpc$^{-1}$\@.
The distance to the Coma cluster is $D_{\rm L}=101$~Mpc, and $1'$
corresponds to 28.9~kpc. We used the abundance ratio by \citet{lod03},
in which the solar Fe abundance relative to H is 2.95$\times$10$^{-5}$.
Errors are quoted at a 90\% confidence level for a single parameter.

The remainder of this paper is organized as follows: 
we present the observations in section 2 and describe the data analysis in section 3. 
We describe the results of spectral fittings and temperature structure in subsection 4.1, 
bulk motions in subsection 4.5, and the Fe abundance in subsection 4.6. 
In section 5, we discuss the results, and in section 6, we summarize the findings.

\section{Observation and Data Reduction}
\label{sec:obs}

\begin{table*}[th]
\caption{Suzaku observation log of the Coma cluster}
\label{tab:ob}
\begin{tabular}{cccccccc} 
\hline
Field name & Target name & Sequence & Observation & RA & Dec & Exposure & Distance from the\\
& & number & date &(deg) & (deg) & (ks) &
X-ray peak\footnotemark[$\ast$] \\ \hline
 \hline
center & Coma Radio Halo & 801097010 & 2006-05-31 & 194.9267 & 27.9061
 & 150 & \timeform{4.7'} \\
\timeform{14'} offset & Coma cluster offset & 801044010 & 2006-05-30 & 194.6939 & 27.9466
& 79 & \timeform{13.9'} \\
\timeform{30'} offset & Coma 11 & 802082010 & 2007-06-19 & 194.6305 & 28.3939 & 53 & \timeform{30.1'} \\
\timeform{34'} offset & Coma BKG2 & 802084010 & 2007-06-21 & 194.3428 & 28.1403 & 30 & \timeform{33.7'} \\
\timeform{44'} offset & Coma 45 & 802047010 & 2007-12-02 & 194.2558 & 27.5714 & 30 & \timeform{44.4'} \\
\timeform{60'} offset & Coma 60 & 802048010 & 2007-12-04 & 194.0251 & 27.4252 & 35 & \timeform{59.5'} \\
 \hline
\timeform{5D} offset & Coma BKG & 802083010 & 2007-06-21 & 198.7472 & 31.6480 & 30 & \timeform{4.9D} \\
\hline
\multicolumn{4}{@{}l@{}}{\hbox to 0pt{\parbox{180mm}{\footnotesize
\footnotemark[$\ast$] X-ray peak has coordinates (RA, Dec) = (194.9367, 27.9472) in degrees.
 }\hss}}
\end{tabular}
\end{table*}

\begin{figure*}[thbp]
\begin{minipage}{0.45\textwidth}
\FigureFile(80mm, 60mm){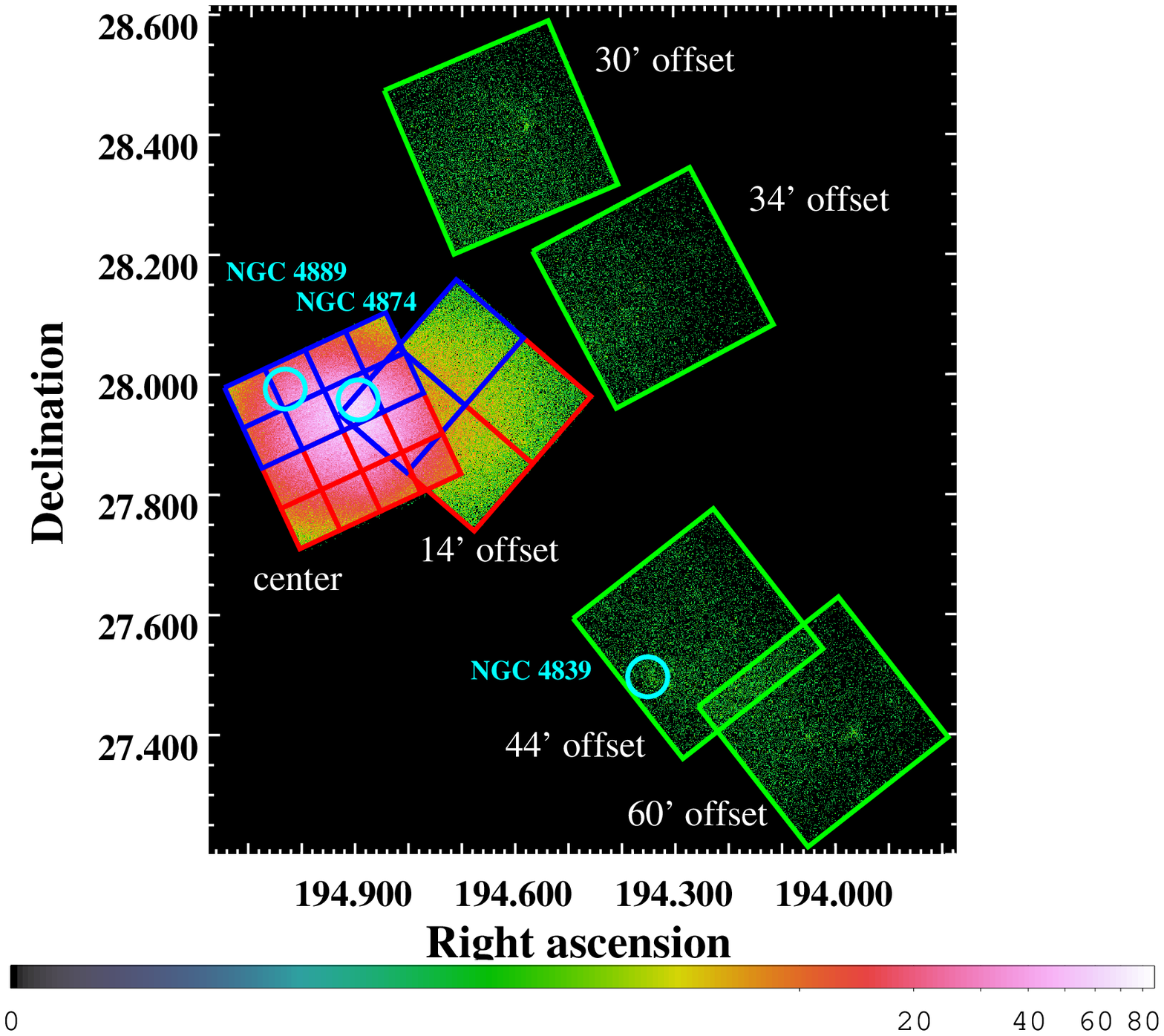}
\end{minipage}\hfill
\begin{minipage}{0.45\textwidth}
\FigureFile(80mm, 60mm){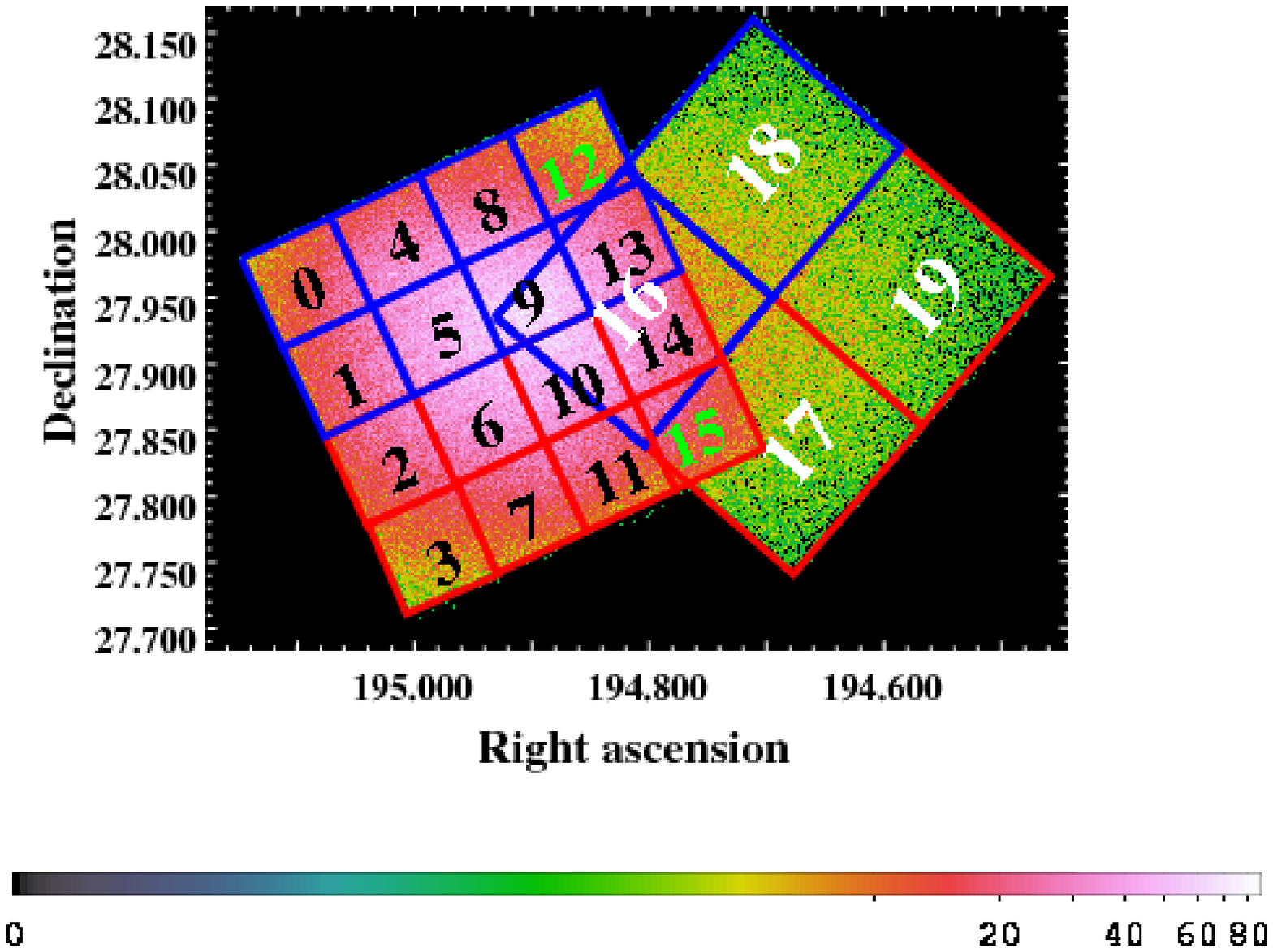}
\end{minipage}\hfill

\caption{XIS0 image (0.5--4.0 keV) of the Coma cluster.
Left: Differences in exposure time and the vignetting effect are uncorrected.
Regions of spectral accumulation are shown as blue, red, and green squares.
The light blue circles show the locations of NGC~4889, NGC~4874, 
and NGC~4839. Right: Definition of the cell numbers in the center and 14$'$ offset regions.
The cell numbers 12 and 15 of the center region (green numbers)
were excluded from our analysis because of strong emissions from calibration sources.}
\label{fig:image}
\end{figure*}

Suzaku carried out six observations in the Coma cluster. The details 
of the observations are summarized in table \ref{tab:ob}. Two central
observations were carried out in May 2006, during the Suzaku 
Phase-I period. The first observation, ``Coma Radio halo'',
has a pointing that is $5'$ offset from the X-ray peak of the Coma cluster,
which has coordinates (RA, Dec) = (194.9367, 27.9472) in degrees.
The second observation, ``Coma cluster offset'', was located 
$14'$ west offset from the first observation. The other observations, 
in the $30'$ and $34'$ north west offset regions and $44'$ and $60'$ south west offset 
regions, were carried out in 2007, during the Suzaku Phase-II period.
To eliminate background emissions, archival Suzaku
data with a 5 degree offset from the Coma cluster, observed in June 2007 
was also used, and is shown in table \ref{tab:ob}.

XIS was operated in its normal mode during observations. 
The XIS instrument consists of four sets of X-ray CCD 
(XIS0, 1, 2, and 3). XIS1 is a back-illuminated (BI) sensor, while 
XIS0, 2, and 3 are front-illuminated (FI) sensors.
XIS2 was not used during the four offset observations beyond 30$'$
from the X-ray peak.
Figure \ref{fig:image} shows a 0.5--4.0 keV image of XIS0.

We performed data reduction using HEAsoft version 6.6.3\@. 
XIS event lists created by the rev~2.0 pipeline processing 
were filtered using the following additional criteria: the geomagnetic cut-off 
rigidity (COR2) $>$ 6 GV, and elevation angle from the earth limb 
$>10^{\circ}$. The data formats of 5x5 and 3x3 editing modes 
were added. The exposure times after the data selection are shown 
in table \ref{tab:ob}. 

To study the temperature structure and search for possible bulk 
motions of ICM in the Coma cluster on the scale of a few arcminutes, 
we divided the 18$'\times18'$ square XIS field of view (FOV) of the center and 14$'$ 
offset regions into $4\times4$ and $2\times2$ cells, respectively,
as shown in figure \ref{fig:image}. 
Each region in the center and $14'$ offset regions subtends 
\timeform{4.5'}$\times$\timeform{4.5'} (117$\times117$ kpc), 
and 9$'\times9'$ (234$\times234$ kpc), respectively. Regions 
around calibration sources were excluded. 
Luminous point sources in which the flux limit corresponds to 
$\sim$ 10$^{-3}$ counts s$^{-1}$ in the 2.0--10.0 keV energy range were also excluded as circular regions 
with a radius of 1$'$. At most 10 point sources were excluded for each of the observed regions.
We extracted spectra from the whole XIS FOV for regions more than $>30'$
from the X-ray peak. The non X-ray background (NXB) was subtracted 
from each spectrum using a database of night Earth observations with 
the same detector area and COR distribution \citep{tawa08}.

We included the degradation of energy resolution due to 
radiation damage in the redistribution matrix file (RMF) generated by 
the {\tt xisrmfgen} Ftools task. We also created an ancillary response 
file (ARF) using the {\tt xissimarfgen} Ftools task \citep{ishisaki07}. 
A decrease in the low-energy transmission of the XIS optical blocking 
filter (OBF) was also included in ARF. To generate ARF 
files for the center and 14$'$ offset regions, we used a
$\beta$-model profile for the simulated surface brightness profile 
in \citet{briel1992}. For $30'$, $34'$, $44'$, $60'$, and 5 degree 
offset regions, we generated ARFs assuming a uniformly extended 
emission from an encircled region with a 20$'$ radius, because the 
gradient of surface brightness within each FOV is small. 
We used the XSPEC\_v11.3.2ag package for spectral analysis.

\section{Data analysis}
\label{sec:analysis}

We have investigated the temperature structure, bulk motions,
and iron abundance distribution of the Coma cluster.
We fitted the spectra with a thermal plasma model (APEC: \cite{smith01}) to obtain the temperature, 
redshift and Fe abundance of ICM. A detailed spectral analysis is presented in subsection \ref{subsec:single}.
We used three Gaussian models 
to obtain the temperature using the normalization ratio
of K$\alpha$ lines of H-like and He-like Fe, and redshift using the shift of line centroid energy,
which is presented in subsection \ref{subsec:ratios}.

\subsection{Estimation of background spectra}

\begin{figure}[htbp]
\begin{center}
\FigureFile(80mm, 60mm){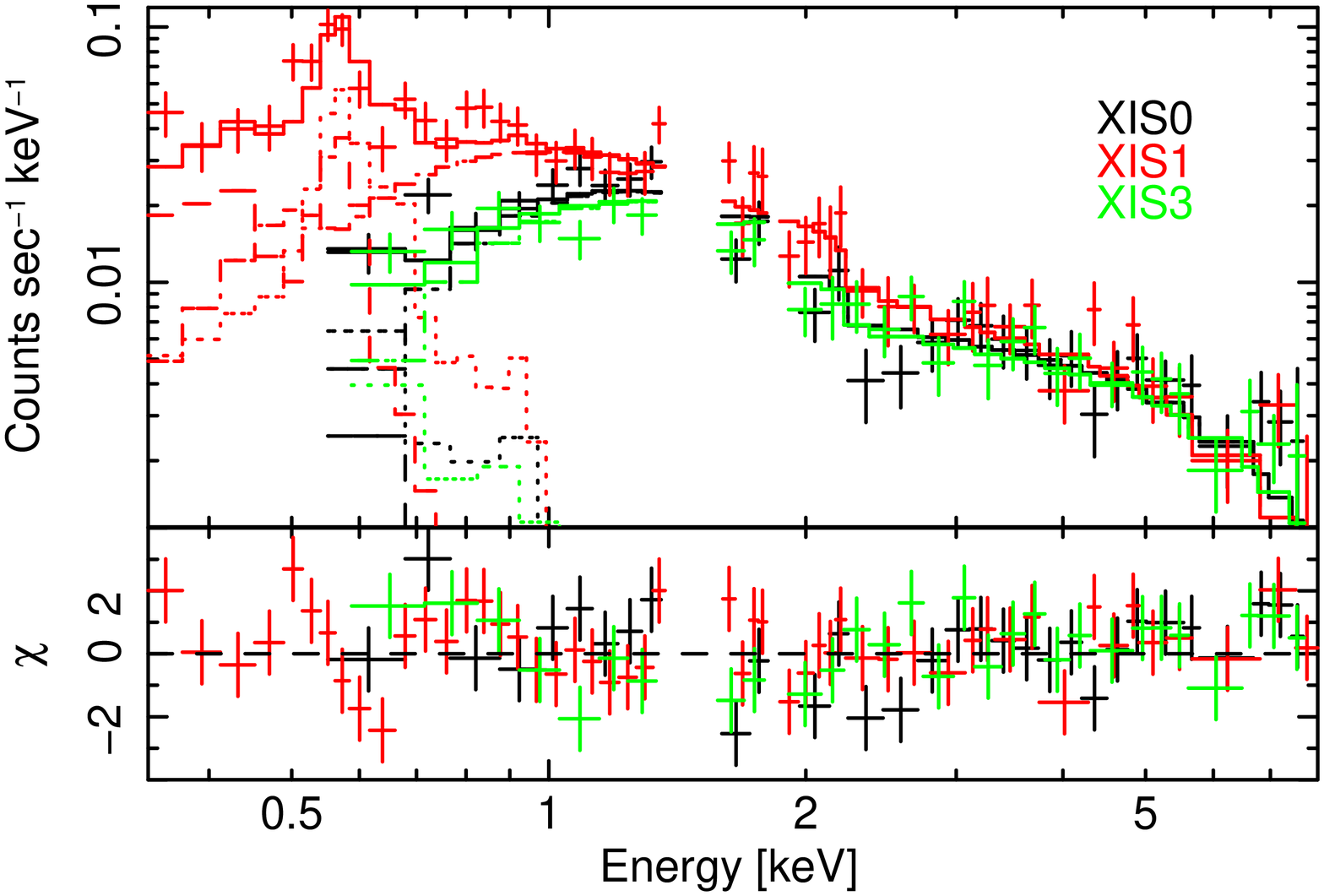}
\end{center}
\caption{NXB-subtracted spectra of the 5 degree offset region in 
the 0.3--8.0 (red: XIS1) and 0.5--8.0 keV energy ranges (black and green for XIS0 and 
3, respectively)\@, ignoring the 1.40--1.55 keV energy range, 
where a relatively large uncertainty exists in the
instrument background \citep{tawa08}. The bottom panel shows the residuals of the fit.}
\label{fig:bkgfit}
\end{figure}

\begin{table*}[tb]
\caption{Resultant parameters of the fits in the background region. 
}
\label{tab:bkgtable}
\begin{tabular}{cccccccc} 
\hline
\multicolumn{2}{c}{Cosmic X-ray Background} & \multicolumn{2}{c}{Local Hot Bubble} & \multicolumn{2}{c}{Milkyway Halo} & \\
\hline \hline
 $\Gamma$ & normalization\footnotemark[$\ast$] & Temperature & normalization\footnotemark[$\dagger$] & Temperature & normalization\footnotemark[$\dagger$] & \multicolumn{2}{c}{Reduced-$\chi^{2}$}  \\
&  & (keV) & $\times 10^{-3}$ & (keV) & $\times 10^{-4}$ & $\chi^2$/d.o.f\footnotemark[$\ddagger$] & $\chi^2$/d.o.f\footnotemark[$\S$]\\
\hline 
$1.38 \pm 0.07$ & $7.6 \pm 0.3$ & 0.07 (fixed) & $6.79 \pm 2.57 $ & $0.16 \pm 0.03$ & $6.55 \pm 2.57$ & 188/141 & 223/154\\
\hline
\multicolumn{4}{@{}l@{}}{\hbox to 0pt{\parbox{180mm}{\footnotesize
\footnotemark[$\ast$] Units: photons cm$^{-2}$ s$^{-1}$ keV$^{-1}$ sr$^{-1}$ at 1 keV.
\par\noindent
\footnotemark[$\dagger$] Normalization of the $APEC$, component divided by the solid angle, $\Omega^{\emissiontype{U}}$, 
assumed in the uniform-sky ARF calculation (\timeform{20'} radius), $Norm = \int n_{e}n_{H}dV/(4\pi(1+z)^{2}D_{A}^{2})/\Omega^{\emissiontype{U}}\times 10^{-14}$ cm$^{-5}$~(400$\pi$)$^{-1}$~arcmin$^{-2}$, where $D_{A}$ is the angular distance to the source.
\par\noindent
\footnotemark[$\ddagger$] Resultant $\chi^2$/d.o.f fitted excluding 1.40--1.55~keV.
\par\noindent
\footnotemark[$\S$] Resultant $\chi^2$/d.o.f fitted including 1.40--1.55~keV.
 }\hss}}
\end{tabular}
\end{table*}

First, we fitted the spectra in the 5$^\circ$ offset region 
to determine the local X-ray background. As shown in \citet{Yoshino2009},
the background emission of Suzaku XIS
can be fitted with a three component model: 
two thermal plasma models (APEC: \cite{smith01}) for the local 
hot bubble (LHB) and the Milky Way halo (MWH), and a power-law model for 
the extragalactic cosmic X-ray background (CXB). MWH and CXB 
components were convolved with an absorption in the Galaxy. 
We, therefore, fitted the spectra simultaneously in the 0.5--8.0 keV energy range for XIS FIs and 
the 0.35--8.0 keV energy range for XIS BI using the following model formula: 
$apec_{\rm LHB}+wabs*(apec_{\rm MWH}+power$-$law_{\rm CXB})$\@.  
We assumed a zero redshift and a solar abundance for LHB and 
MWH components. The temperature of LHB was fixed at 0.07 keV 
\citep{takei08}, while the temperature of MWH was variable. 
The column density of the Galactic neutral hydrogen was fixed at
1.0$\times10^{20}$ cm$^{-2}$ \citep{kalberla05}. 
To avoid systematic uncertainties in the background,
we ignored the 1.40--1.55 keV energy range, where a relatively large uncertainty exists  
because of an instrument Al line \citep{tawa08}, and energies above 8 keV.
We also excluded the narrow energy 
band between 1.82 and 1.84 keV in the fits because of incomplete 
responses around the Si edge.
Results of the spectral fit are shown in figure \ref{fig:bkgfit} 
and the resultant parameters of 
the fit are shown in table \ref{tab:bkgtable}. 
The derived photon index and normalization 
of the CXB component agree with \citet{kushino02} and \citet{takei08}.
The derived temperature and normalizations of the two thermal models 
are consistent with those obtained from XMM and Suzaku observations
\citep{Lumb2002, Yoshino2009}.

\begin{figure*}[ht]
\FigureFile(80mm, 60mm){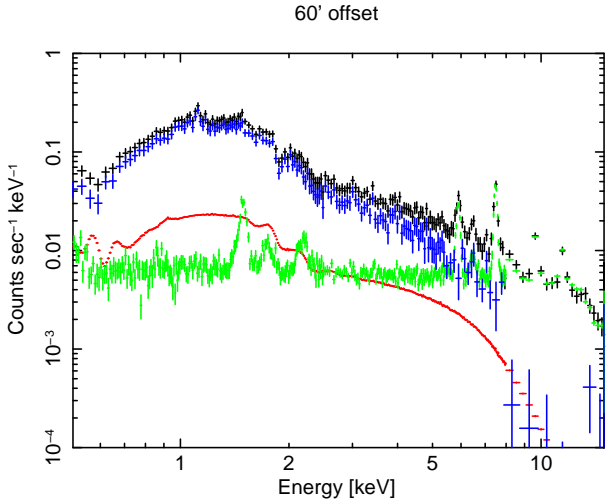}
\caption{XIS0 spectra of the 60$'$ offset region of 
the Coma cluster (black), NXB (green), and
the simulated background spectrum including LHB, MWH and CXB (red). 
 Blue crosses show the background-subtracted spectrum of the 60$'$ offset region of the Coma cluster.
}
\label{fig:bgd}
\end{figure*}

We generated the simulated spectra of the X-ray background
including LHB, MWH, and CXB 
emissions derived from the resultant parameters of the fits in 
the 5$^\circ$ offset region. We used the simulated spectra as 
the background to fit the spectra for ICM of the Coma cluster.  
Figure \ref{fig:bgd} shows the comparison between the XIS0
spectra of the 60$'$ offset region of the Coma cluster and backgrounds. 
Above 7 keV, NXB dominates the observed spectra of the 60$'$ offset region.
We also show the background-subtracted spectrum of the 60$'$ offset region of 
the Coma cluster in figure \ref{fig:bgd} (blue).
The X-ray background spectra are well reproduced.

\subsection{Spectral fits with single-, two- and three temperature models}
\label{subsec:single}

\begin{figure*}[htbp]

\begin{minipage}{0.45\textwidth}
\FigureFile(80mm, 60mm){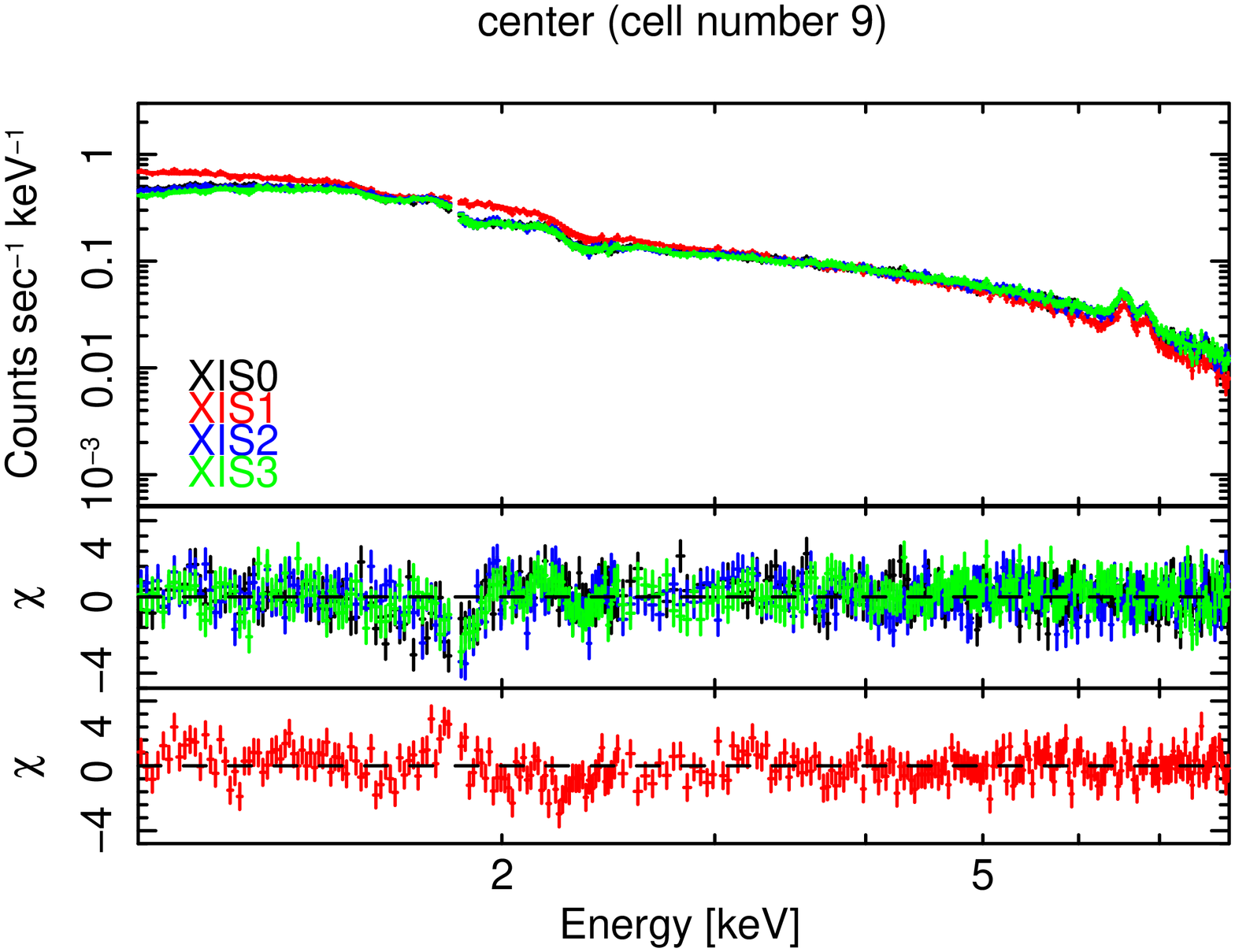}
\end{minipage}\hfill
\begin{minipage}{0.45\textwidth}
\FigureFile(80mm, 60mm){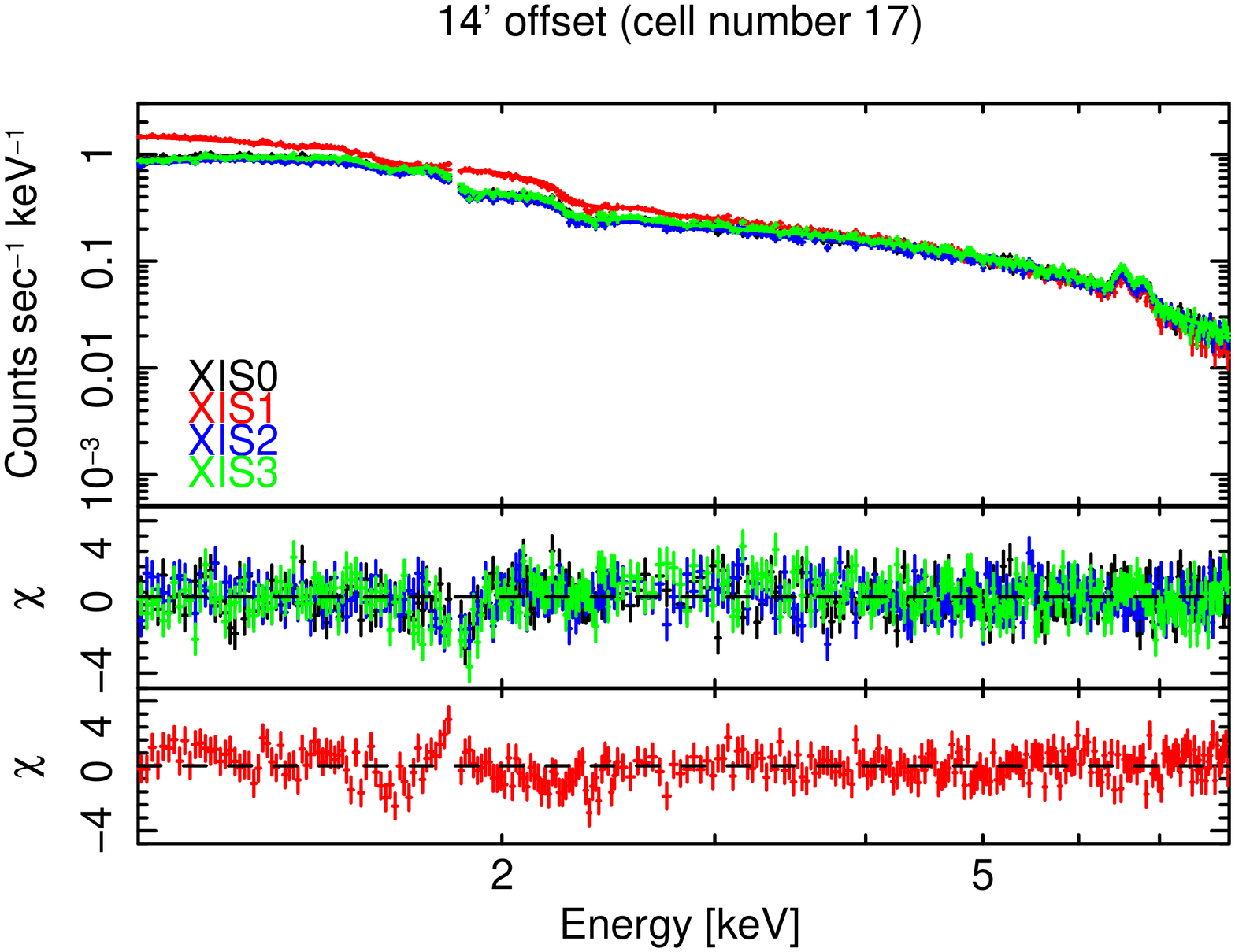}
\end{minipage}\hfill

\begin{minipage}{0.45\textwidth}
\FigureFile(80mm, 60mm){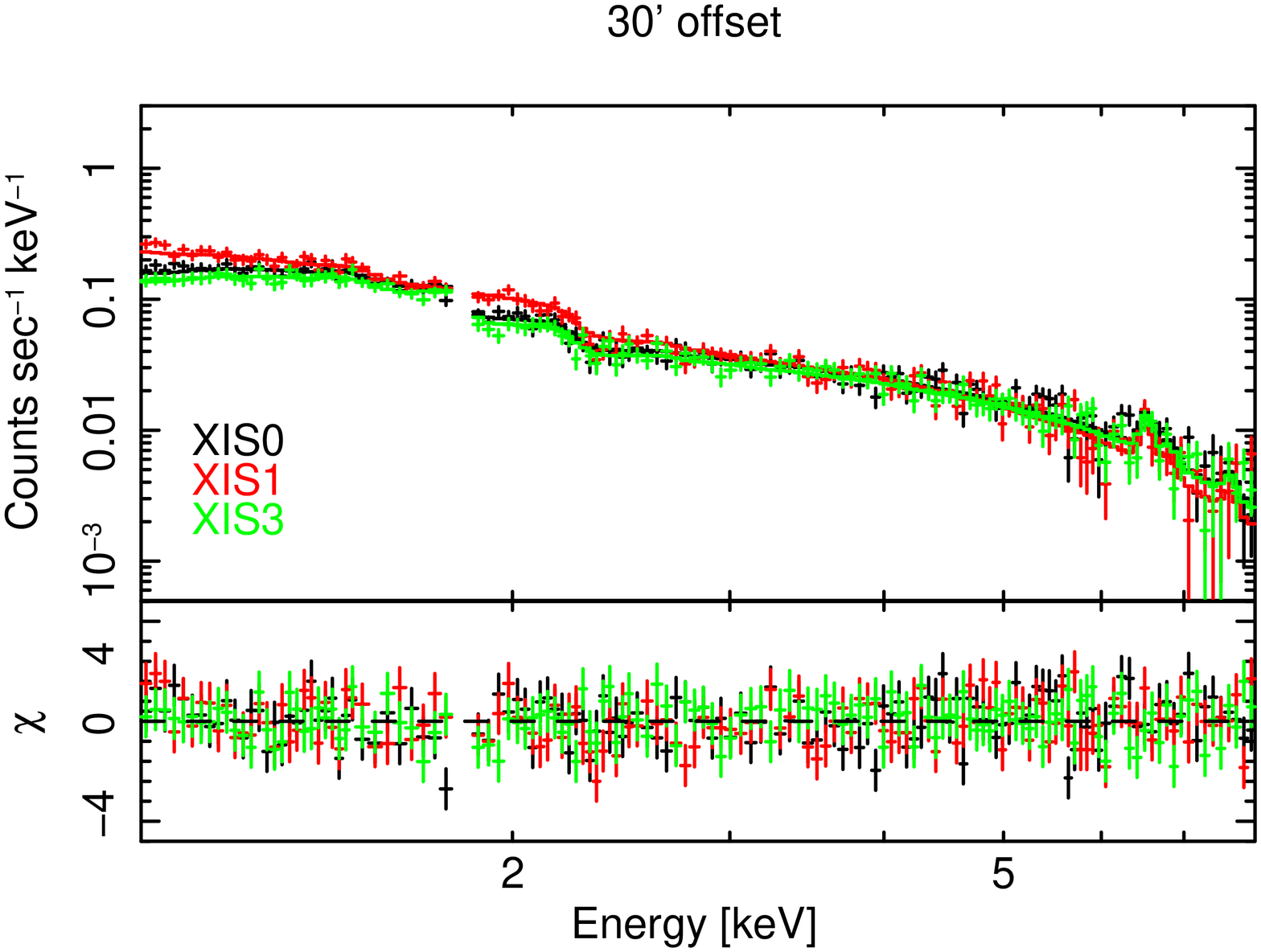}
\end{minipage}\hfill
\begin{minipage}{0.45\textwidth}
\FigureFile(80mm, 60mm){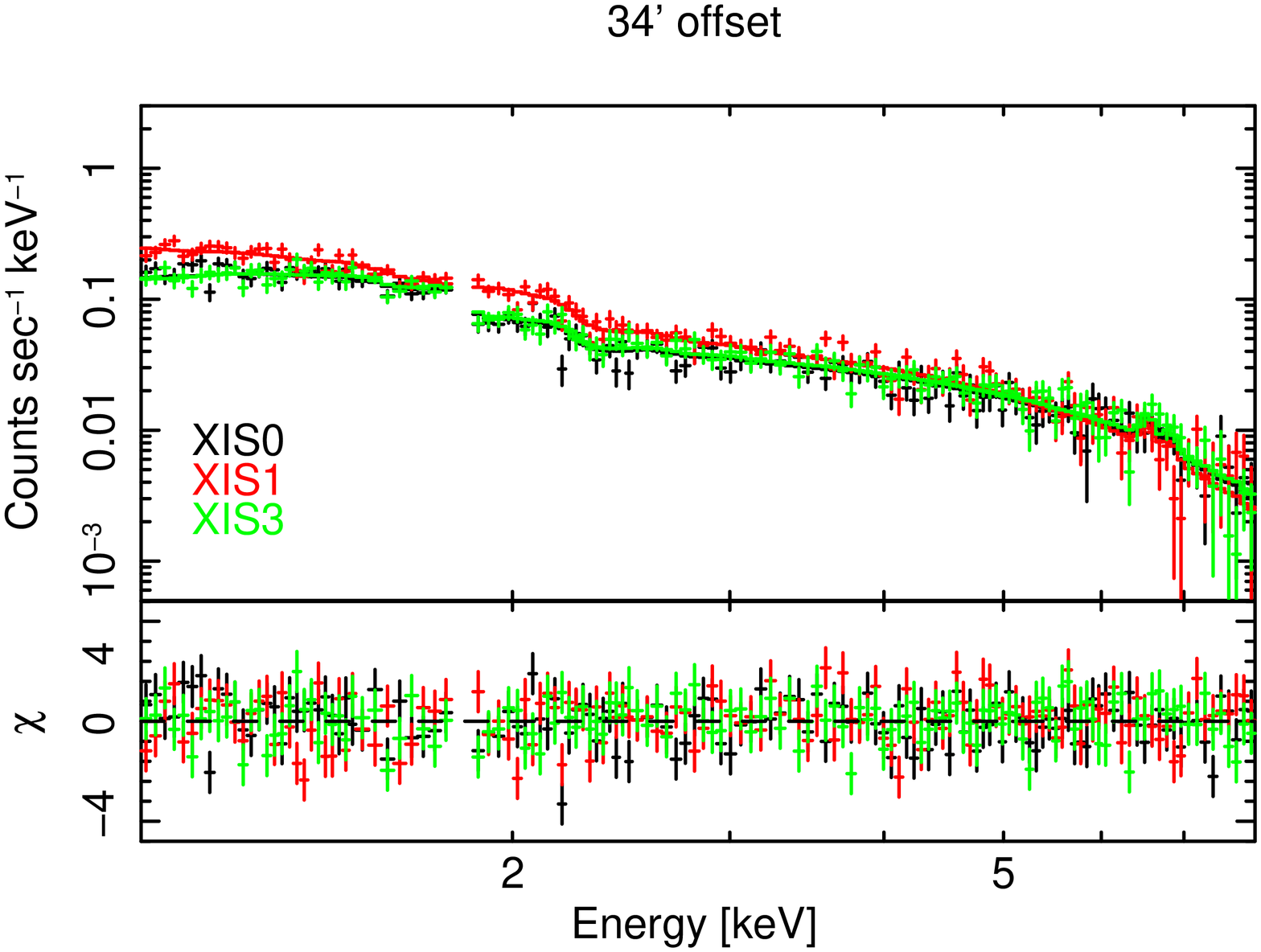}
\end{minipage}\hfill

\begin{minipage}{0.45\textwidth}
\FigureFile(80mm, 60mm){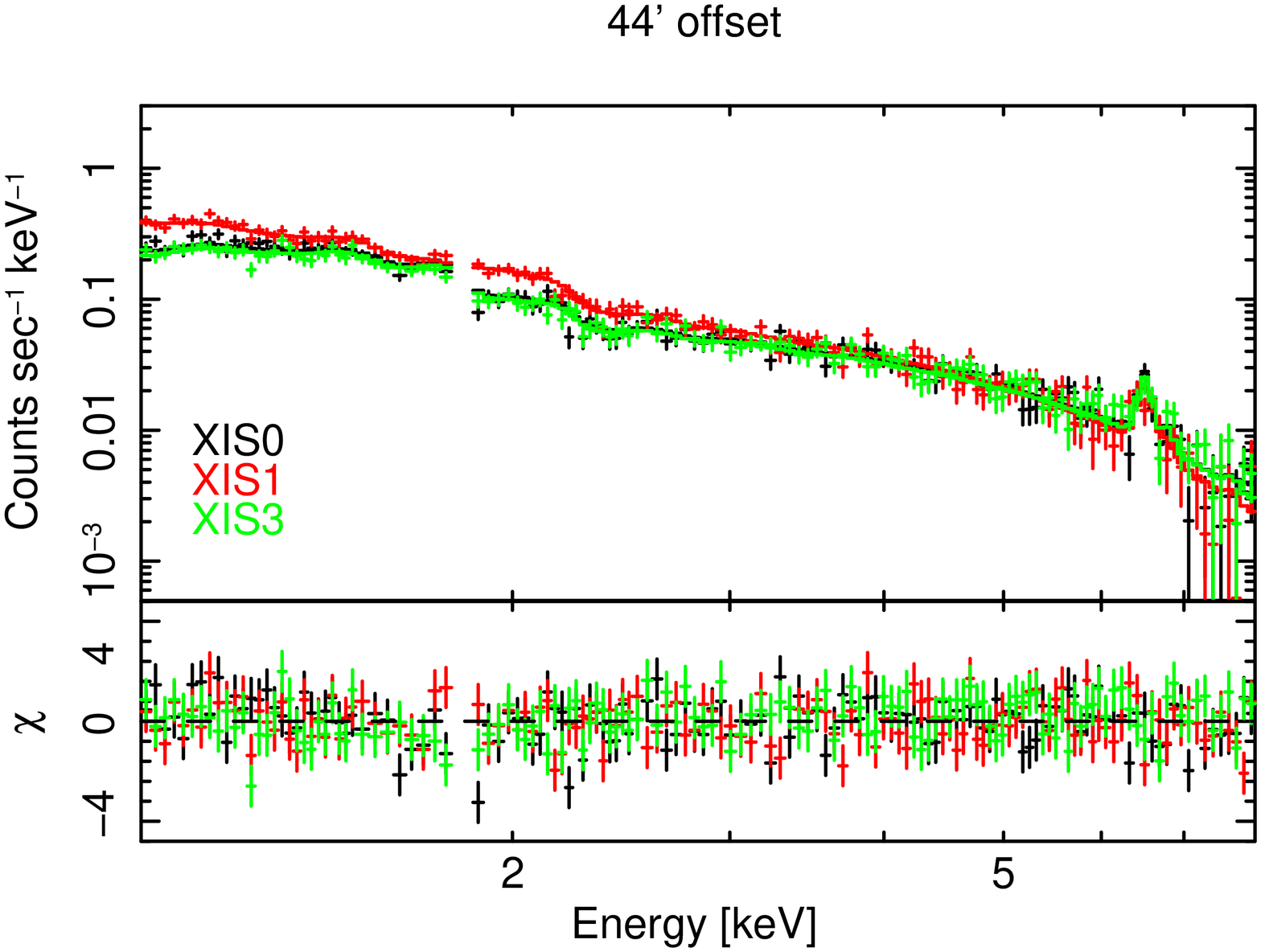}
\end{minipage}\hfill
\begin{minipage}{0.45\textwidth}
\FigureFile(80mm, 60mm){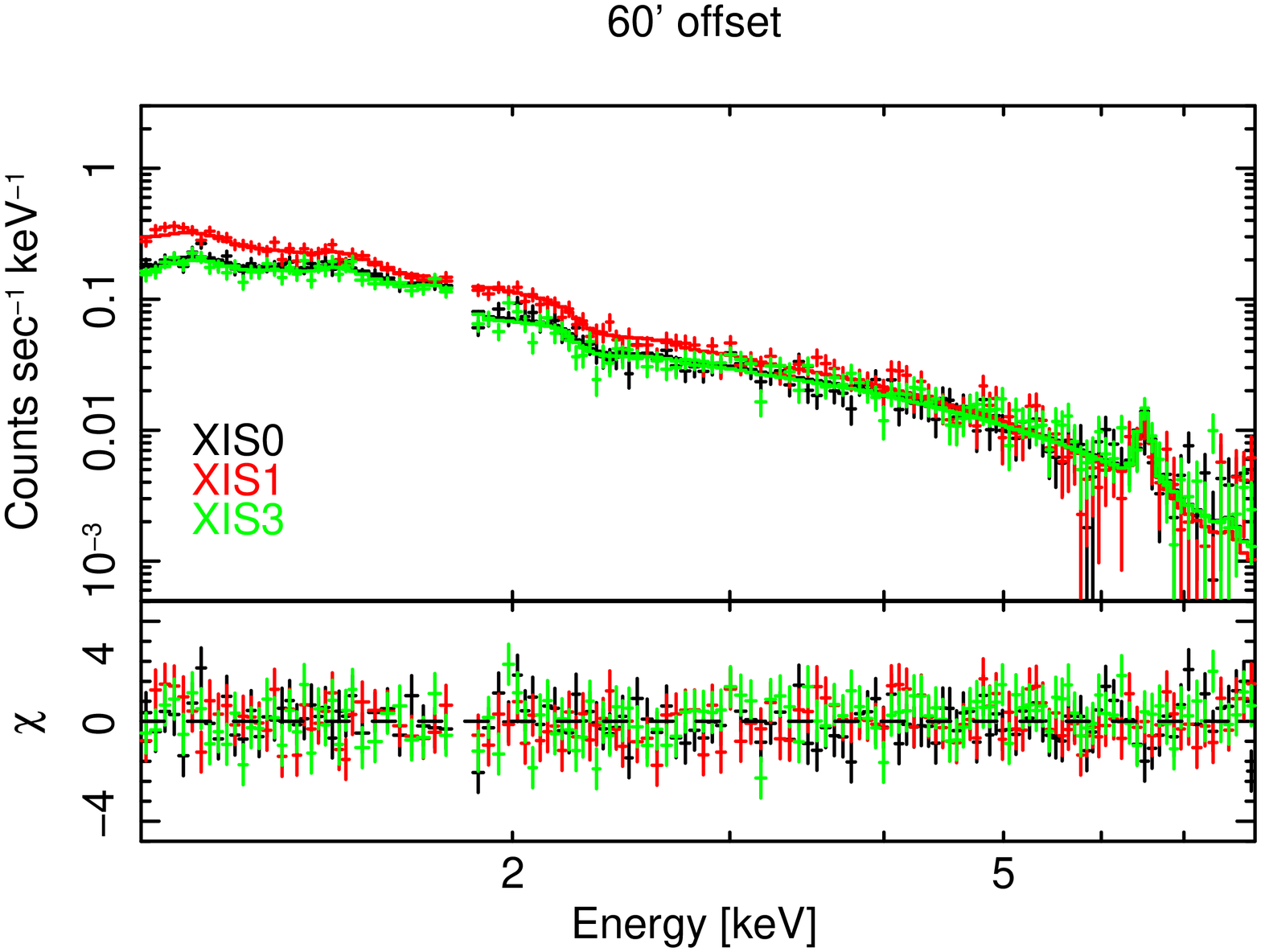}
\end{minipage}\hfill

\caption{Spectra for one cell in the center and 14$'$ offset regions, 
and all other offset regions, 
fitted by the single temperature (APEC) model.
Black, red, blue, and green crosses and lines correspond to the 
XIS0, 1, 2, and 3, respectively.
The middle and bottom panels show the residuals of the fit. 
}
\label{fig:spec}
\end{figure*}

We fitted the XIS spectra in each region of the Coma cluster using a single temperature 
 model (APEC) with the Galactic absorption, $N_{\rm H}$\@.
Each spectral bin contained 50 or more counts.  
We fitted the spectra in two energy bands: 1.0--8.0 and 5.0--8.0 keV\@.
To avoid systematic uncertainties in the background, 
we ignored energies higher than 8 keV.
We also excluded the narrow energy 
band between 1.82 and 1.84 keV in the fits because of incomplete 
responses around the Si edge. Temperature, redshift and normalization 
of the single temperature  model were free parameters, and 
$N_{\rm H}$ was fixed to the Galactic value, 
$1.0\times 10^{20}~{\rm  cm^{-2}}$, in the direction of the Coma 
cluster. Abundances of He, C, N, and Al were fixed to be 
a solar. We divided the other metals into three groups: O, Ne and Mg; 
Si, S, Ar and Ca; and Fe and Ni, and allowed them to vary. 
The resultant parameters are summarized in table~
\ref{tab:1-8vapecresults}, and a sample of the spectra is shown in 
figure~\ref{fig:spec}.
Since our data have very high statistics, especially in the central region,
there remain residual structures around the Si edge structure.
These residuals are probably caused by systematic uncertainties in the response
matrix because the BI and FI detectors gave different residuals.
Therefore, to avoid systematic uncertainties in all regions, 
we excluded the energy range of 1.7--2.3 keV and fitted the spectra again.
The derived temperatures hardly changed, although the reduced 
$\chi^2$ decreased by 20\%. The resultant $\chi^2$ are summarized in table
\ref{tab:1-8vapecresults}.

To constrain the temperature structure, 
we fitted the spectra of the center and offset regions 
with a two-temperature (APEC) model in the energy range of 1.0--8.0 keV,
ignoring the energy range of 1.7--2.3 keV.
The $\chi^{2}$ are improved slightly by
less than a few percent.
The derived temperatures of one component are similar to those derived from the
single-temperature model fits, and those of the other component are
either above 10 keV or below 2 keV.
To obtain more detailed constraints for the lower and higher temperature components,
we fitted the spectra with a three-temperature (APEC) model in the energy range of 1.0--8.0 keV,
excluding the 1.7--2.3 keV energy range to avoid uncertainties in the response matrix. 
In this model, the lowest, middle and highest temperatures are restricted to below 6 keV,
within 6.0--10.0 keV, and above 10.0 keV, respectively.
The resultant $\chi^{2}$ values are shown in table \ref{tab:1-8vapecresults}.

To derive the Fe abundance and velocity of the bulk-motion from Fe-K lines,
we fitted the spectra of each region with the single-temperature APEC model
in an energy range of 5.0--8.0 keV.
The results are shown in table \ref{tab:1-8vapecresults}.

To precisely determine the bulk velocity of ICM,
the accuracy of photon energy measurements is crucial.
Some sensors have large uncertainties in the determination of the line centroid energy. 
Thus, we averaged the spectrum over FOV in (i) all detectors, and (ii) all detectors 
except for different Mn K$\alpha$ line centroid energies to check the spectral shape and measured redshift 
in the center and NGC~4839 regions (44$'$ offset region). 

\small
\tabcolsep 4pt
\begin{table*}[th]
\caption{Resultant parameters of the fits with the single-temperature (APEC) model 
in the energy ranges of 1.0--8.0 keV and 5.0--8.0 keV, the two- or three-temperature model, 
and three Gaussians models\@.
In the case of three Gaussian models, we converted the normalization ratio
of K$\alpha$ lines of H-like and He-like Fe to temperature with the APEC model.}
\label{tab:1-8vapecresults}
\begin{tabular}{cccccccccc}
\hline \hline
& \multicolumn{5}{c}{APEC fit in 1.0--8.0 keV} & \multicolumn{3}{c}{APEC fit in 5.0--8.0 keV} & Gaussians\\
\hline
 Region & $kT$ & $\chi^{2}$/d.o.f.\footnotemark[$\ast$] & $\chi^{2}$/d.o.f.\footnotemark[$\dagger$] & $\chi^{2}$/d.o.f.\footnotemark[$\ddagger$]& $\chi^{2}$/d.o.f.\footnotemark[$\S$] & Fe & Redshift & $\chi^{2}$/d.o.f.  & $kT$ from line\\
 & [keV]  & (1T) & (1T) & (2T) & (3T) & [solar] & [$10^{-2}$] &  & ratio
 [keV]\\
\hline
center & & & & & & & & &\\
$0$ & $8.47_{-0.08}^{+0.14}$ & $938/829$ & $776/711$ & $772/709$ & $770/707$ & $0.47_{-0.05}^{+0.06}$ & $2.17_{-0.16}^{+0.16}$ & $265/244$  & $8.08_{-0.80}^{+0.80}$ \\
$1$ & $7.81_{-0.10}^{+0.09}$ & $1373/1108$ & $1037/950$ & $999/948$ & $996/946$ & $0.40_{-0.04}^{+0.02}$ & $2.24_{-0.10}^{+0.09}$ & $319/328$ & $7.75_{-0.48}^{+0.51}$\\ 
$2$ & $7.19_{-0.09}^{+0.11}$ & $1325/1108$ & $1015/950$ & $1007/948$ & $1004/946$ & $0.39_{-0.03}^{+0.03}$ & $2.09_{-0.10}^{+0.11}$ & $344/328$ & $6.86_{-0.52}^{+0.52}$\\
$3$ & $7.12_{-0.19}^{+0.18}$ & $901/829$ & $728/711$ & $728/709$ & $726/707$ & $0.33_{-0.09}^{+0.04}$ & $2.11_{-0.09}^{+0.19}$ & $245/244$ &  $6.78_{-0.91}^{+0.94}$\\
$4$ & $8.73_{-0.16}^{+0.12}$ & $1363/1108$ & $1082/950$ & $1075/948$ & $1075/946$ & $0.37_{-0.03}^{+0.03}$ & $2.42_{-0.12}^{+0.09}$ & $352/328$  & $8.61_{-0.54}^{+0.56}$\\
$5$ & $8.20_{-0.07}^{+0.07}$ & $1574/1108$ & $1105/950$ & $1082/948$ & $1081/946$ & $0.40_{-0.02}^{+0.03}$ & $2.25_{-0.08}^{+0.07}$ & $394/328$  & $8.25_{-0.37}^{+0.38}$\\
$6$ & $7.81_{-0.07}^{+0.08}$ & $1298/1108$ & $994/950$ & $962/948$ & $961/946$ & $0.42_{-0.03}^{+0.02}$ & $2.14_{-0.07}^{+0.08}$ & $346/328$ &  $8.17_{-0.40}^{+0.40}$\\
$7$ & $7.45_{-0.12}^{+0.11}$ & $1314/1108$ & $1018/950$ & $975/948$ & $971/946$ & $0.48_{-0.05}^{+0.04}$ & $2.23_{-0.13}^{+0.19}$ & $333/328$ &  $7.70_{-0.67}^{+0.63}$\\
$8$ & $8.93_{-0.15}^{+0.15}$ & $1290/1108$ & $994/950$ & $993/948$ & $993/946$ & $0.38_{-0.04}^{+0.04}$ & $2.42_{-0.10}^{+0.14}$ & $390/328$ &  $9.45_{-0.63}^{+0.66}$\\ 
$9$ & $8.66_{-0.12}^{+0.09}$ & $1533/1108$ & $1049/950$ & $1000/948$ & $999/946$ & $0.40_{-0.02}^{+0.01}$ & $2.33_{-0.02}^{+0.08}$ & $318/328$ &  $8.93_{-0.40}^{+0.41}$\\
$10$ & $8.26_{-0.07}^{+0.07}$ & $1395/1108$ & $1005/950$ & $975/948$ & $968/946$ & $0.38_{-0.03}^{+0.02}$ & $2.14_{-0.08}^{+0.07}$ & $349/328$ &  $8.31_{-0.41}^{+0.41}$\\
$11$ & $8.15_{-0.13}^{+0.13}$ & $1379/1108$ & $1068/950$ & $1041/948$ & $1041/946$ & $0.37_{-0.04}^{+0.04}$ & $2.21_{-0.12}^{+0.15}$ & $369/328$ &  $8.23_{-0.65}^{+0.65}$\\
$13$ & $9.08_{-0.23}^{+0.15}$ & $1361/1108$ & $1106/950$ & $1103/948$ & $1103/946$ & $0.33_{-0.03}^{+0.04}$ & $2.33_{-0.12}^{+0.13}$ & $378/328$ &  $9.23_{-0.74}^{+0.66}$\\
$14$ & $8.95_{-0.16}^{+0.15}$ & $1306/1108$ & $1046/950$ & $1029/948$ & $1029/946$ & $0.39_{-0.04}^{+0.04}$ & $2.31_{-0.14}^{+0.15}$ & $357/328$ &  $8.96_{-0.71}^{+0.71}$\\
\hline
\timeform{14'}offset & & & & & & & & &\\
$16$ & $9.14_{-0.14}^{+0.13}$ & $1355/1108$ & $1047/950$ & $1035/948$ & $1035/946$ & $0.40_{-0.03}^{+0.03}$ & $2.38_{-0.07}^{+0.08}$ & $352/328$ &  $8.83_{-0.54}^{+0.54}$\\
$17$ & $9.22_{-0.10}^{+0.19}$ & $1201/1108$ & $987/950$ & $974/948$ & $967/946$ & $0.34_{-0.05}^{+0.04}$ & $2.21_{-0.08}^{+0.18}$ & $332/328$ &  $9.29_{-0.89}^{+0.92}$\\
$18$ & $9.00_{-0.22}^{+0.19}$ & $1303/1108$ & $1087/950$ & $1078/948$ & $1077/946$ & $0.36_{-0.07}^{+0.05}$ & $2.39_{-0.24}^{+0.13}$ & $396/328$ & $9.22_{-0.89}^{+0.89}$\\
$19$ & $8.43_{-0.17}^{+0.16}$ & $1280/1108$ & $1067/950$ & $1054/948$ & $1054/946$ & $0.27_{-0.05}^{+0.05}$ & $2.23_{-0.25}^{+0.22}$ & $389/328$ & $7.82_{-1.17}^{+1.18}$\\
\hline
\timeform{30'} offset & $6.65_{-0.25}^{+0.29}$ & $452/397$ & $387/343$ &
 $370/341$ & $368/339$ & $0.26_{-0.13}^{+0.06}$ & $2.30_{-0.43}^{+0.45}$ & $135/94$ & $7.29_{-2.76}^{+2.74}$\\
\timeform{34'} offset & $8.46_{-0.46}^{+0.67}$ & $484/397$ & $408/343$ &
 $405/341$ & $405/339$ & $0.21_{-0.10}^{+0.13}$ & $1.92_{-1.07}^{+1.10}$ & $107/94$ & $7.67_{-1.72}^{+1.74}$\\ 
\timeform{44'} offset & $5.31_{-0.15}^{+0.20}$ & $445/397$ & $360/343$ &
 $350/341$ & $349/339$ & $0.37_{-0.08}^{+0.09}$ & $2.69_{-0.30}^{+0.33}$ & $106/94$ & $6.33_{-1.63}^{+1.64}$\\
\timeform{60'} offset & $3.74_{-0.13}^{+0.13}$ & $401/397$ & $344/343$ &
 $344/341$ & $325/339$ & $0.29_{-0.10}^{+0.10}$ & $2.67_{-0.51}^{+0.50}$ & $93/94$ & $2.19(<5.95)$\\
\hline
\multicolumn{4}{@{}l@{}}{\hbox to 0pt{\parbox{180mm}{\footnotesize
\footnotemark[$\ast$] $\chi^{2}$/d.o.f. of single-temperature spectral fits including
the 1.7--2.3 keV energy range.
\par\noindent
\footnotemark[$\dagger$] $\chi^{2}$/d.o.f. of single-temperature spectral fits excluding
the 1.7--2.3 keV energy range.
\par\noindent
\footnotemark[$\ddagger$] $\chi^{2}$/d.o.f. of two-temperature
 spectral fits excluding the 1.7--2.3 keV energy range.
\par\noindent
\footnotemark[$\S$] $\chi^{2}$/d.o.f. of three-temperature spectral fits excluding 
the 1.7--2.3 keV energy range.
 }\hss}}
\end{tabular}
\end{table*}

\subsection{Central energy and normalization ratio of Fe lines}
\label{subsec:ratios}

\begin{figure*}[htbp]
\begin{minipage}{0.45\textwidth}
\FigureFile(80mm, 60mm){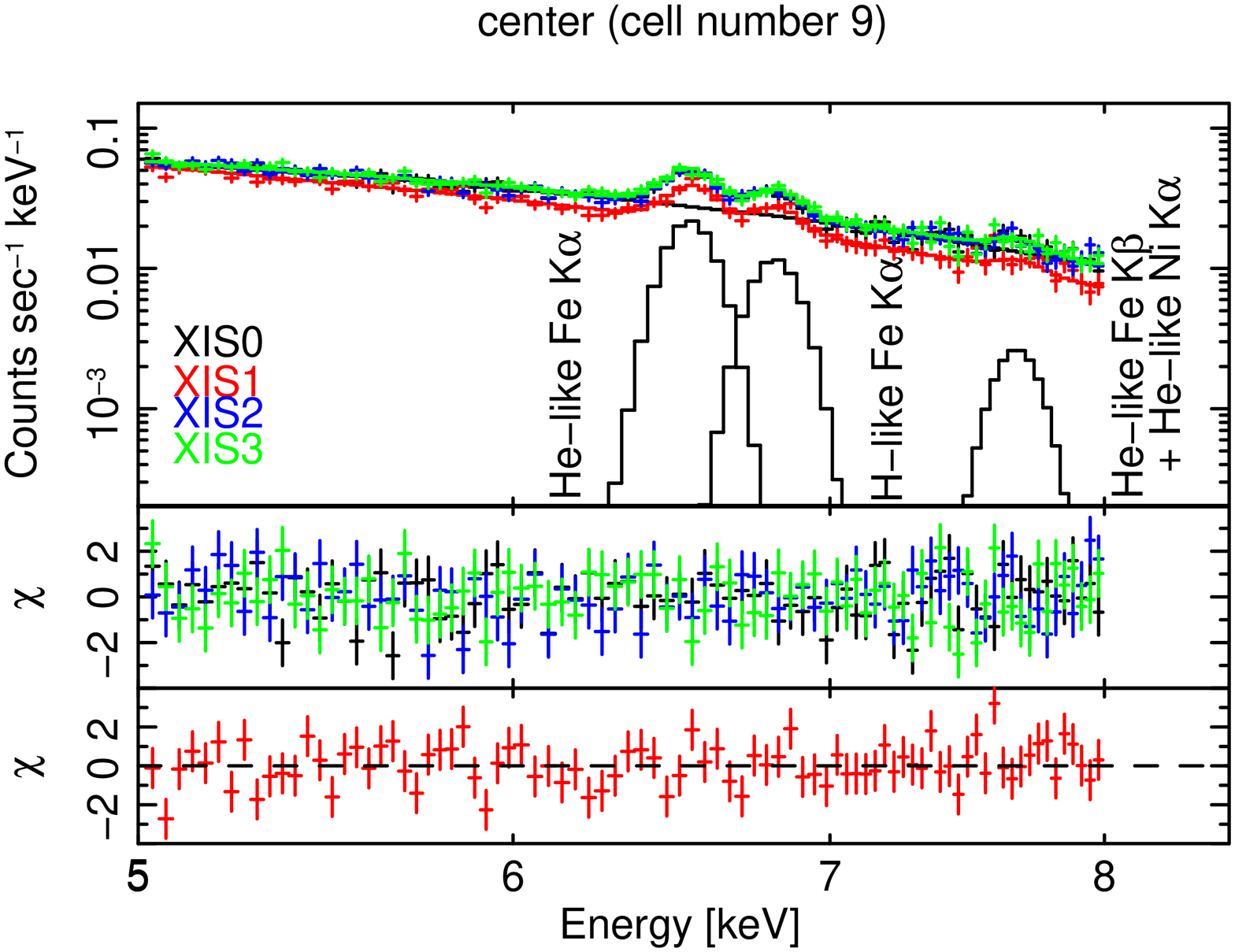}
\end{minipage}\hfill
\begin{minipage}{0.45\textwidth}
\FigureFile(80mm, 60mm){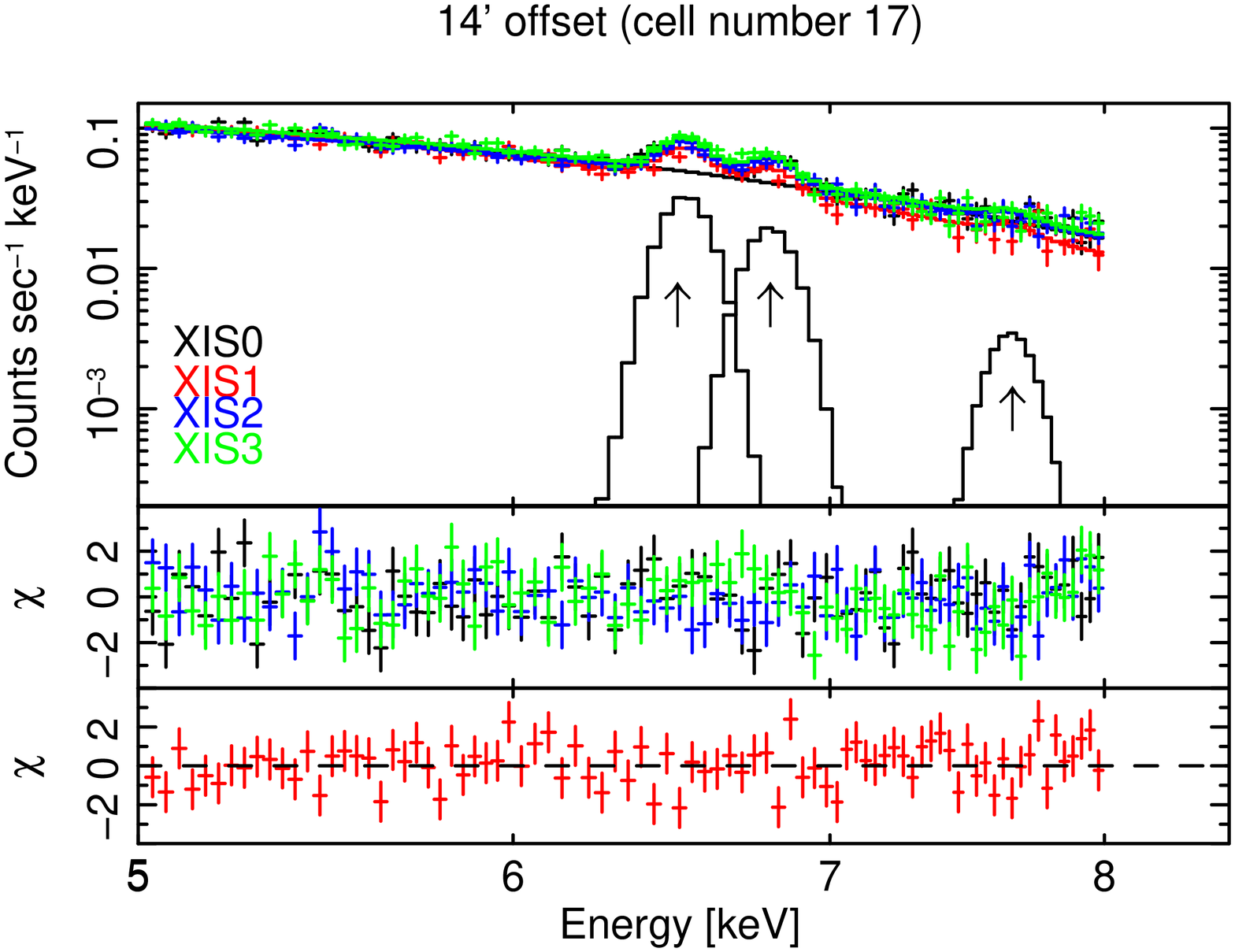}
\end{minipage}

\begin{minipage}{0.45\textwidth}
\FigureFile(80mm, 60mm){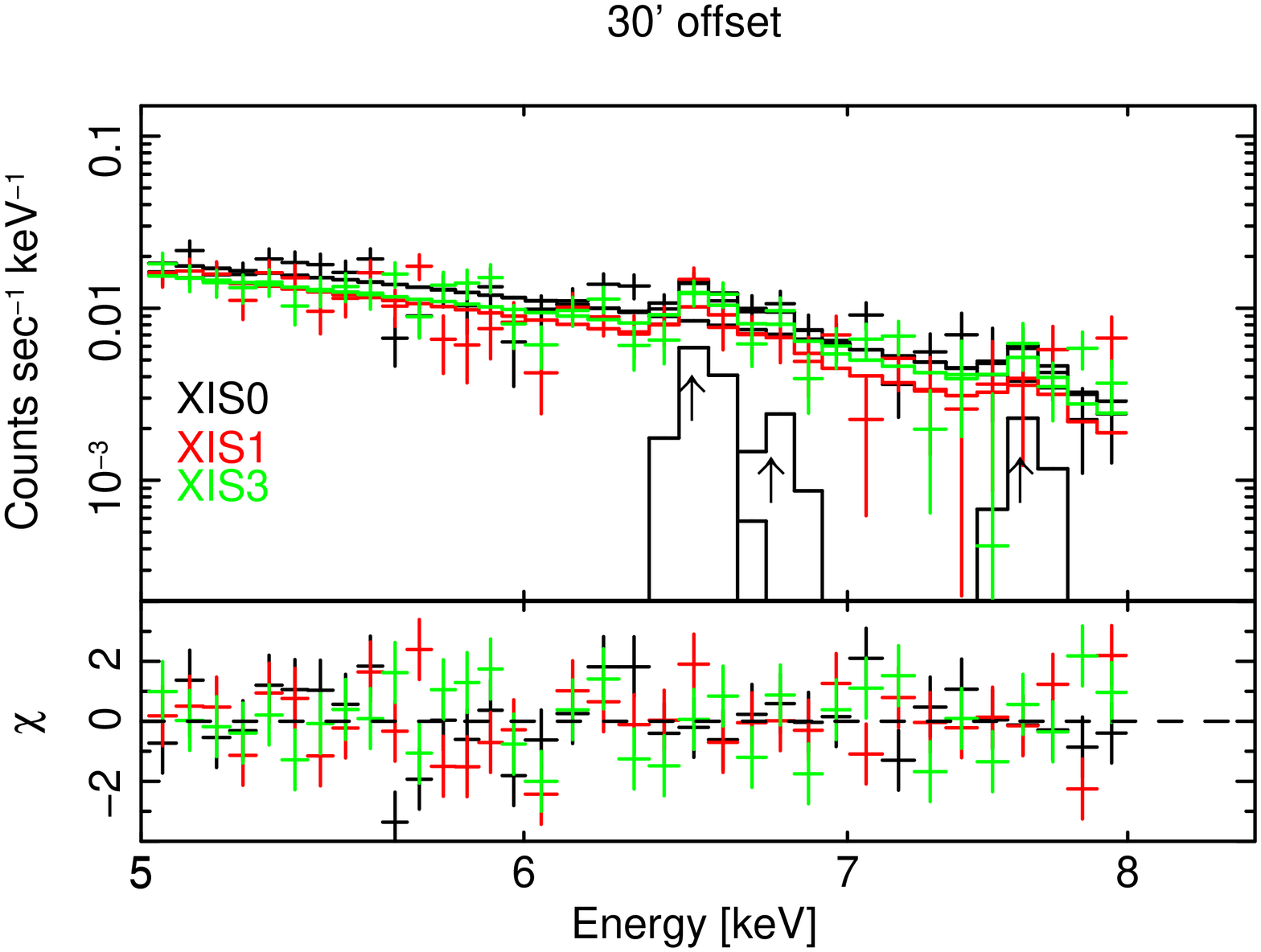}
\end{minipage}\hfill
\begin{minipage}{0.45\textwidth}
\FigureFile(80mm, 60mm){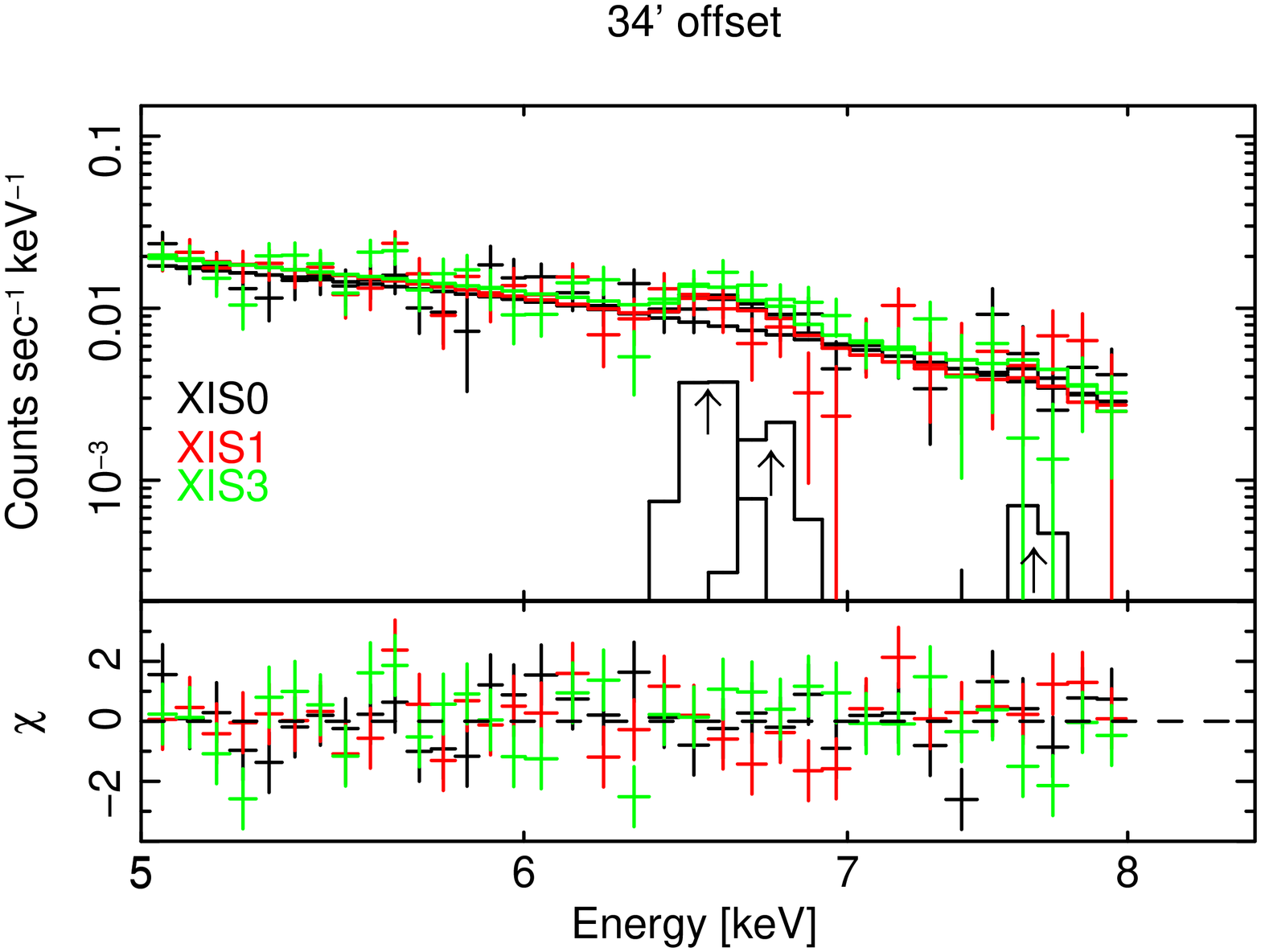}
\end{minipage}

\begin{minipage}{0.45\textwidth}
\FigureFile(80mm, 60mm){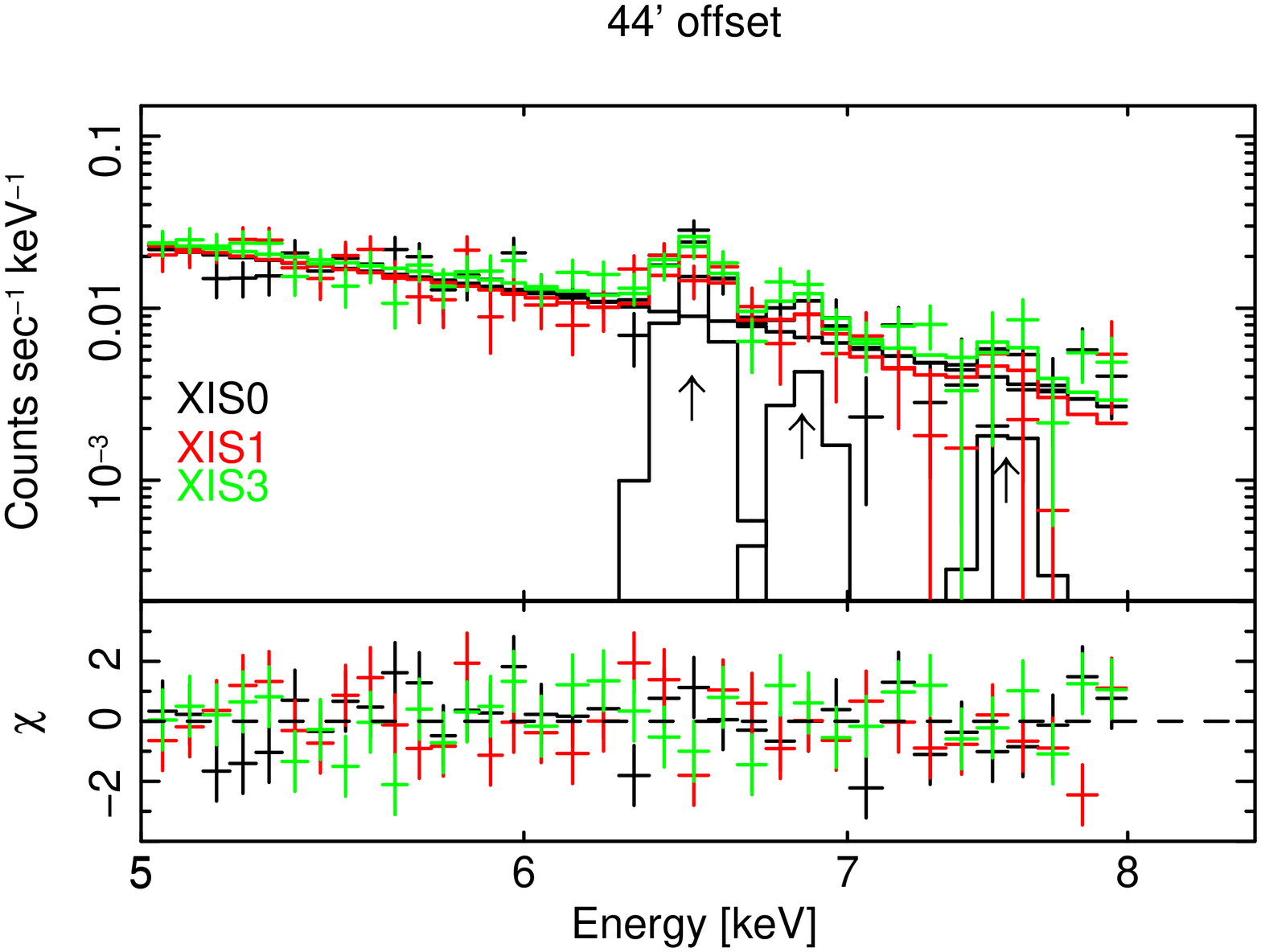}
\end{minipage}\hfill
\begin{minipage}{0.45\textwidth}
\FigureFile(80mm, 60mm){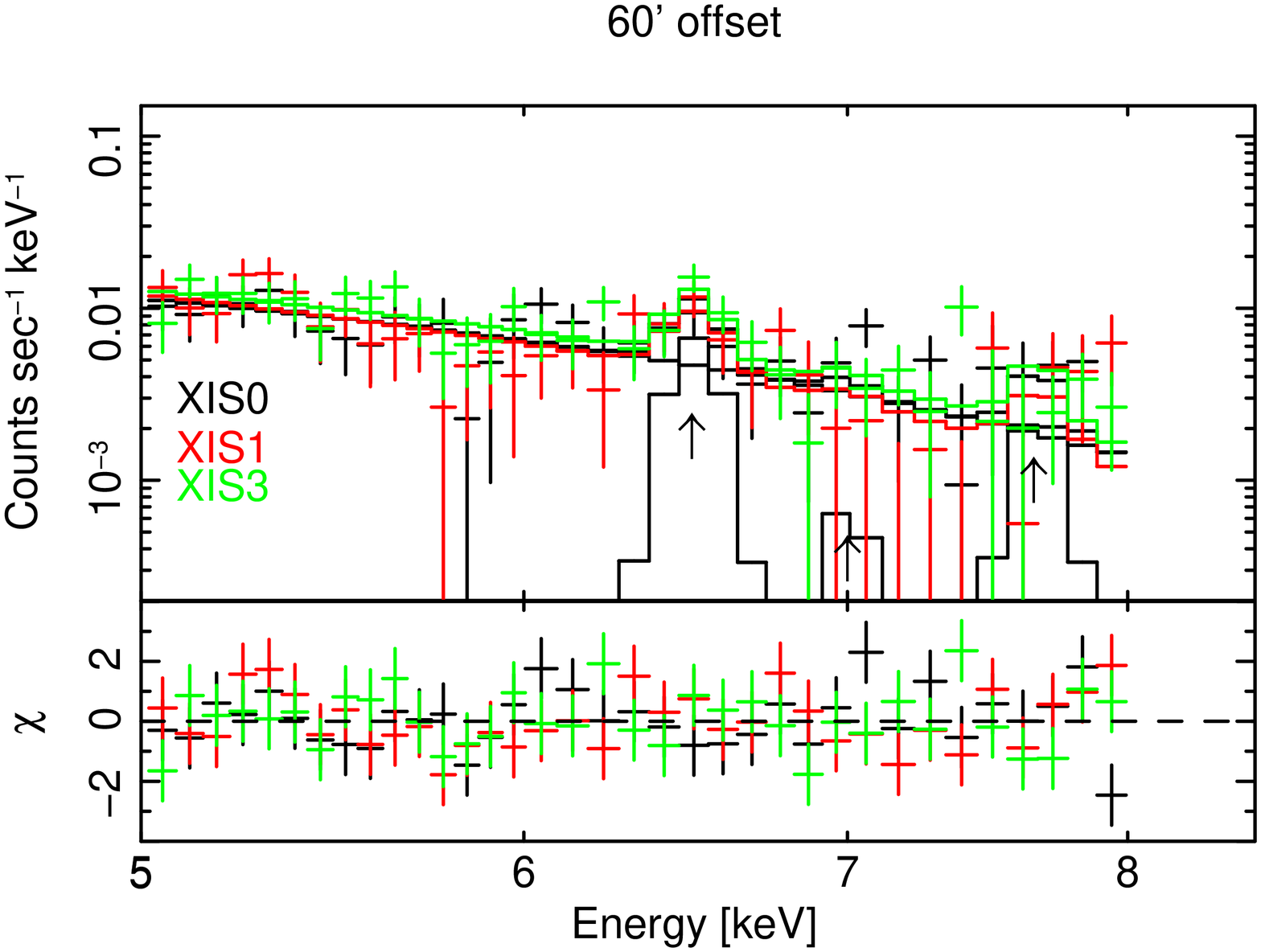}
\end{minipage}

\caption{
Representative iron K$\alpha$ line spectra (for the same regions as in figure \ref{fig:spec}),
fitted with the bremsstrahlung and three Gaussian model in the energy range of 
5.0--8.0 keV\@. Three Gaussian models (black lines) represent the strong 
K$\alpha$ line of He-like Fe, a weaker line of H-like Fe, 
and a mixture of the K$\alpha$ line of He-like Ni and the K$\beta$ line of 
He-like Fe. 
}
\label{fig:bremss_spec}
\end{figure*}

To derive the ICM temperature from the normalization ratio
of K$\alpha$ lines of H-like and He-like Fe, and the possible bulk 
motions of ICM from the central energy of the lines, we fitted 
the spectra in the energy range of 5.0--8.0 keV with a sum of bremsstrahlung and 
three Gaussian components, which correspond to K$\alpha$ lines of He-like 
and H-like Fe, and a mixture of the K$\alpha$ line of He-like Ni and the K$\beta$ 
line of He-like Fe. The temperature and normalization of the 
bremsstrahlung component, and the central energy and normalizations of 
the Gaussian components were free parameters, while the line width of 
Gaussian models was fixed at 0\@.
The best-fit spectra are shown in figure \ref{fig:bremss_spec}.

\section{Results}

\subsection{Temperature structure of ICM}
\label{subsec:temp}

\begin{figure*}[htbp]
\begin{minipage}{0.45\textwidth}
\FigureFile(80mm,60mm){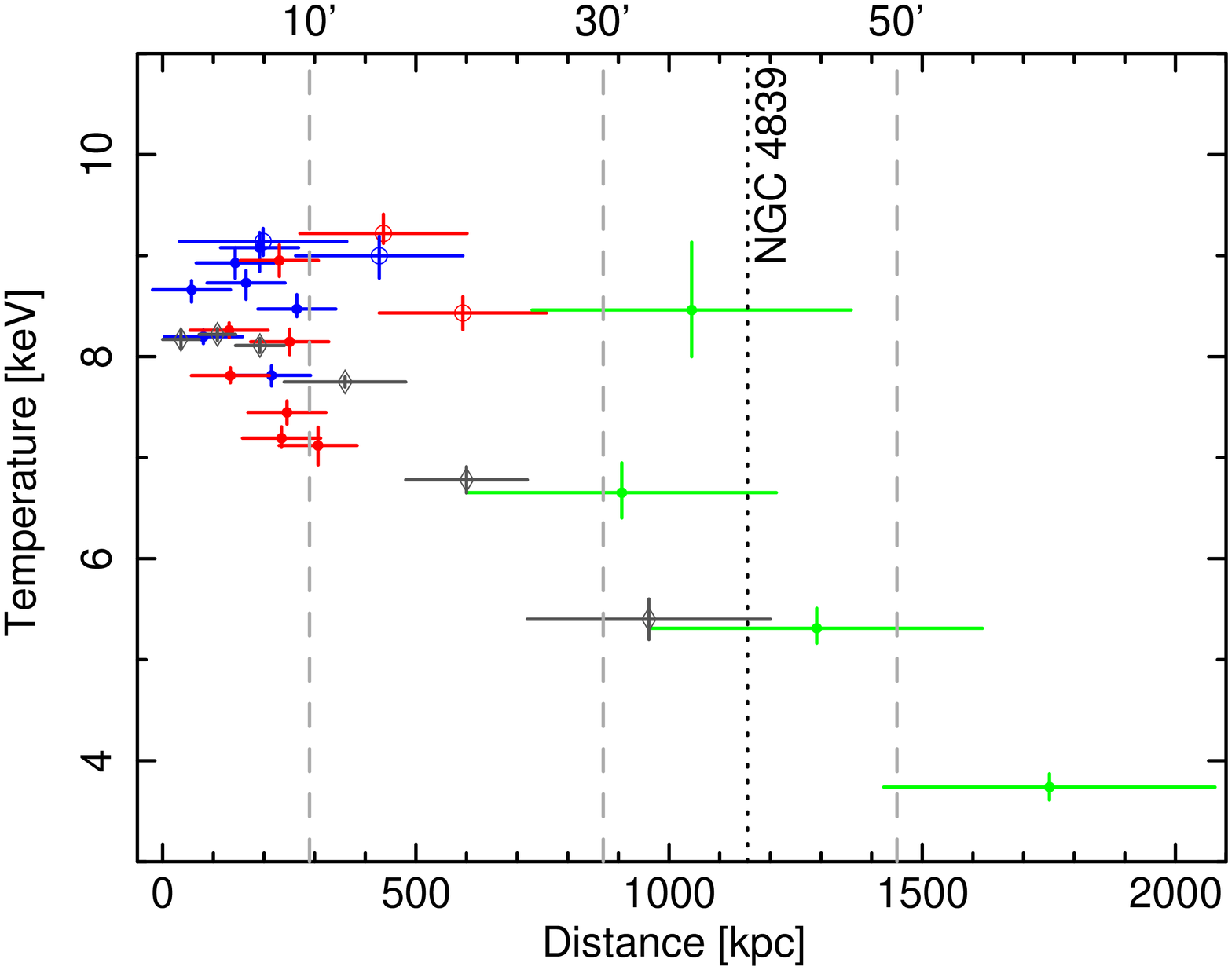}
\end{minipage}\hfill
\begin{minipage}{0.45\textwidth}
\FigureFile(80mm,60mm){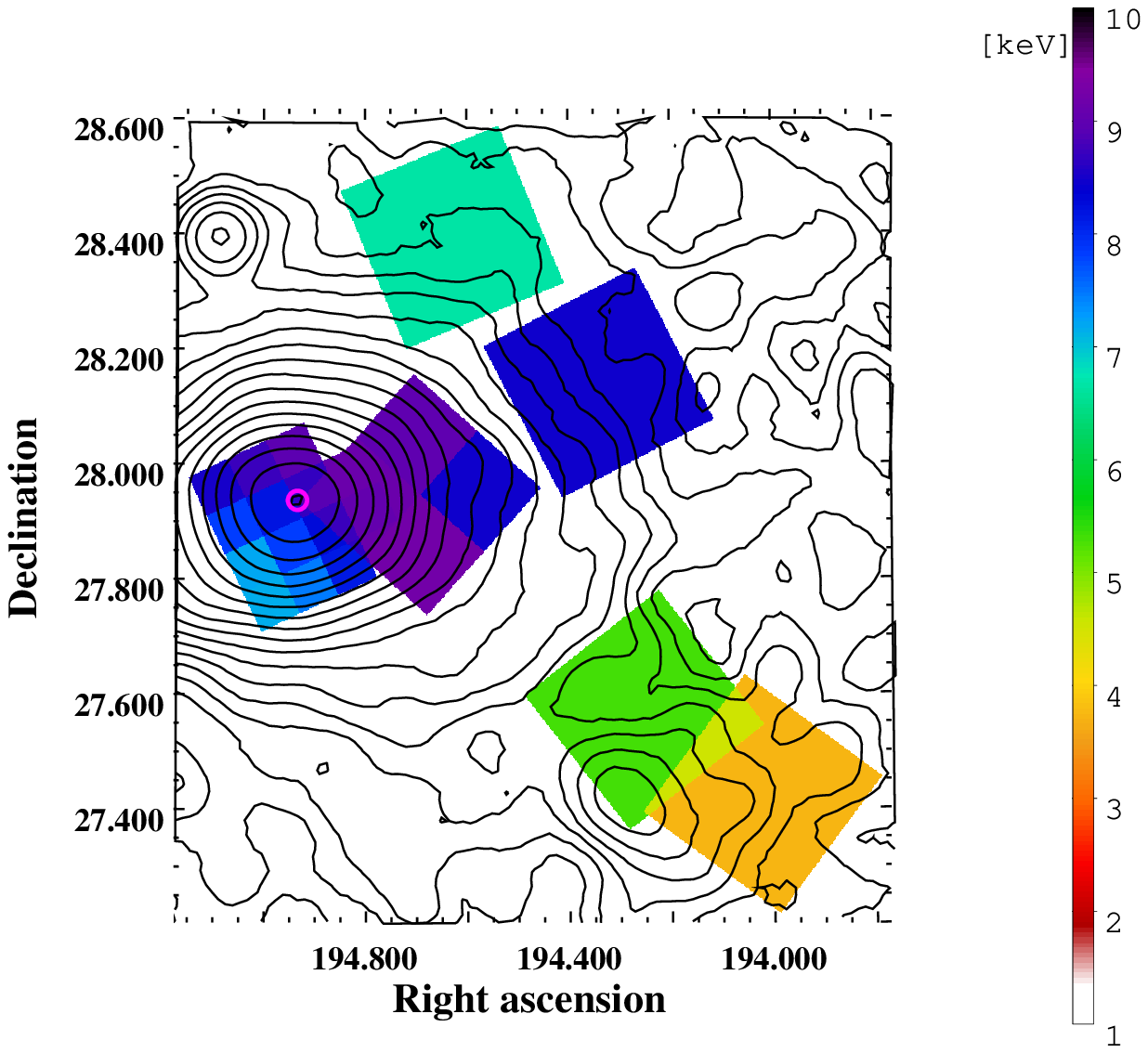}
\end{minipage}\hfill

\caption{Radial temperature profile from the X-ray peak (left) 
and temperature map (right) from our observations of the Coma cluster. 
Temperatures are derived from the spectral fits with the 
single temperature (APEC) model in the energy range of 1.0--8.0 keV\@. The color notations 
in the left panel are same as those in figure \ref{fig:image}.
Clear and shaded circles correspond to the center 
and 14$'$ offset regions, respectively, and gray diamonds indicate 
the results from XMM-Newton \citep{Matsushita2011}. 
The magenta circle in the right panel shows the X-ray peak of the 
Coma cluster. The temperature map is overlaid on 
X-ray contours on a linear scale in the 0.4--2.4 keV energy range from the ROSAT-All-Sky-Survey.
}
\label{fig:kT_apec}
\end{figure*}

Figure \ref{fig:kT_apec} shows a radial profile and map of 
the derived temperatures from the spectral fits with the 
single temperature (APEC) model in the energy range of 1.0--8.0 keV\@. The radial 
temperature profile of the Coma cluster is relatively flat within 
$\sim$500 kpc, which corresponds to the 14$'$ offset region, 
and decreases with the radius to the outer region. While the regions 
observed with Suzaku are limited, the resultant temperature 
profile is similar to that derived from XMM-Newton 
observations, which cover nearly the entire Coma cluster out 
to $\sim$1200 kpc \citep{Matsushita2011}.

Excluding the south-east quadrant (cell numbers: 2,3,6, and 7), and the 14$'$ offset region, 
temperatures of cells in the center region 
have a flat distribution within a small range 
between $\sim$8 keV and $\sim$9 keV\@. Within 300 kpc from the 
X-ray peak in the center region, the weighted mean temperature with statistical errors 
of the 16 cells is 8.34$\pm$0.19 keV\@.
The south-east quadrant (cell numbers: 2,3,6, and 7) in the center region 
has a slightly cooler temperature: the weighted average with statistical errors of the 
four cells is 7.39$\pm$0.11 keV\@. This value is close to 
that of 5.0--7.0 keV in the cool temperature region observed by 
XMM-Newton, which would be a counterpart of the south-east quadrant 
region \citep{arnaud01}\@.
On the other hand, the north-west quadrant in the center region 
(cell numbers: 8,13, and 14) and the nearby area of the 14$'$ offset region 
(cell numbers: 17,18) have slightly higher temperatures of $\sim$9 keV\@. 
The 34$'$ offset region located to the north-west of 
the 14$'$ offset region, also has a relatively high temperature of 
$\sim$8--9 keV\@. This direction corresponds to the hot spot, 
$kT=12.7_{-2.0}^{+3.6}$ keV in \citet{donnelly1999} with ASCA, 
and $kT=8.4\pm$0.4 keV in \citet{arnaud01} with XMM-Newton. 
Recent INTEGRAL observations have shown a surface brightness excess 
relative to XMM observations, which is well-represented by the 
extended hot thermal emission with $\sim12\pm2$ keV \citep{eckert2007}\@.
The temperature in the 30$'$ offset region is around $\sim$6--7 keV,
and is clearly cooler than temperatures in the inner region.
This value is consistent with the previous results in
\citet{honda1996}, and \citet{takei08}.

The 44$'$ and 60$'$ offset regions include the NGC~4839 subcluster,
and temperatures decreases with radius to 5.31$\pm$0.20 keV 
and 3.74$\pm$0.13 keV, respectively.
These values are cooler than those in \citet{honda1996} with ASCA, 
but are consistent with those in \citet{wik09} and \cite{neumann2001} 
with XMM-Newton. Temperatures of the core and tail of NGC~4839 
with XMM-Newton, which is located in the 44$'$ offset region with Suzaku, 
are $3.1_{-2.4}^{+3.7}$ keV and $4.8_{-4.0}^{+6.0}$ keV\@,
respectively \citep{neumann2001}. The temperature of the main body 
of the subcluster is $4.4_{-4.7}^{+4.0}$ keV \citep{neumann2001}\@.
Suzaku provided more significant temperatures than previous measurements
in these regions.

In summary, the temperature distribution observed with
Suzaku is fairly consistent with the previous results observed with ASCA,
XMM and INTEGRAL. 
In addition, we note that statistical and systematic 
errors with Suzaku are much smaller than those in previous results.

\subsection{Fits with three-temperature models}

\begin{figure*}[htbp]

\begin{minipage}{0.45\textwidth}
\FigureFile(80mm,60mm){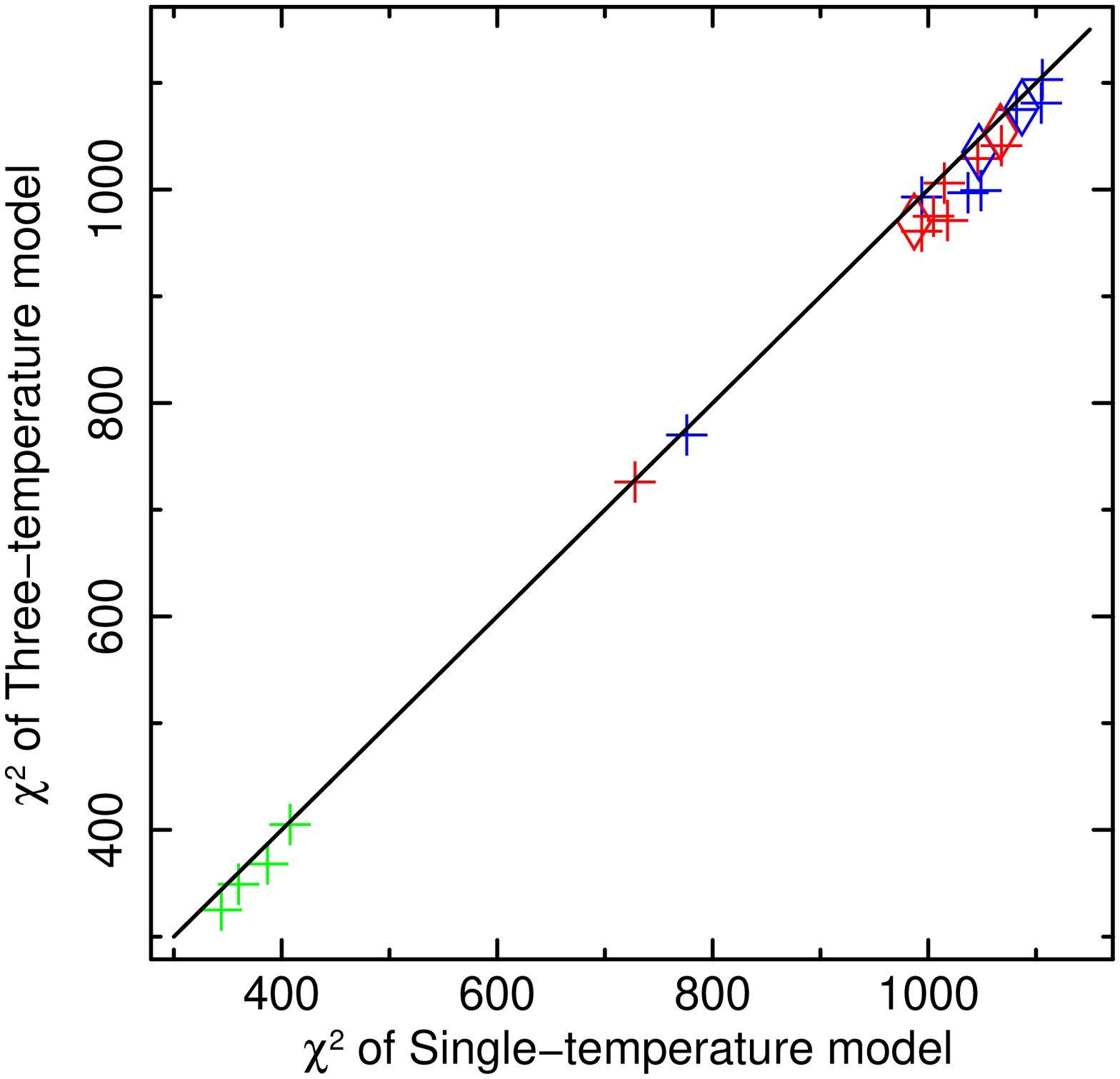}
\end{minipage}\hfill
\begin{minipage}{0.45\textwidth}
\FigureFile(80mm,60mm){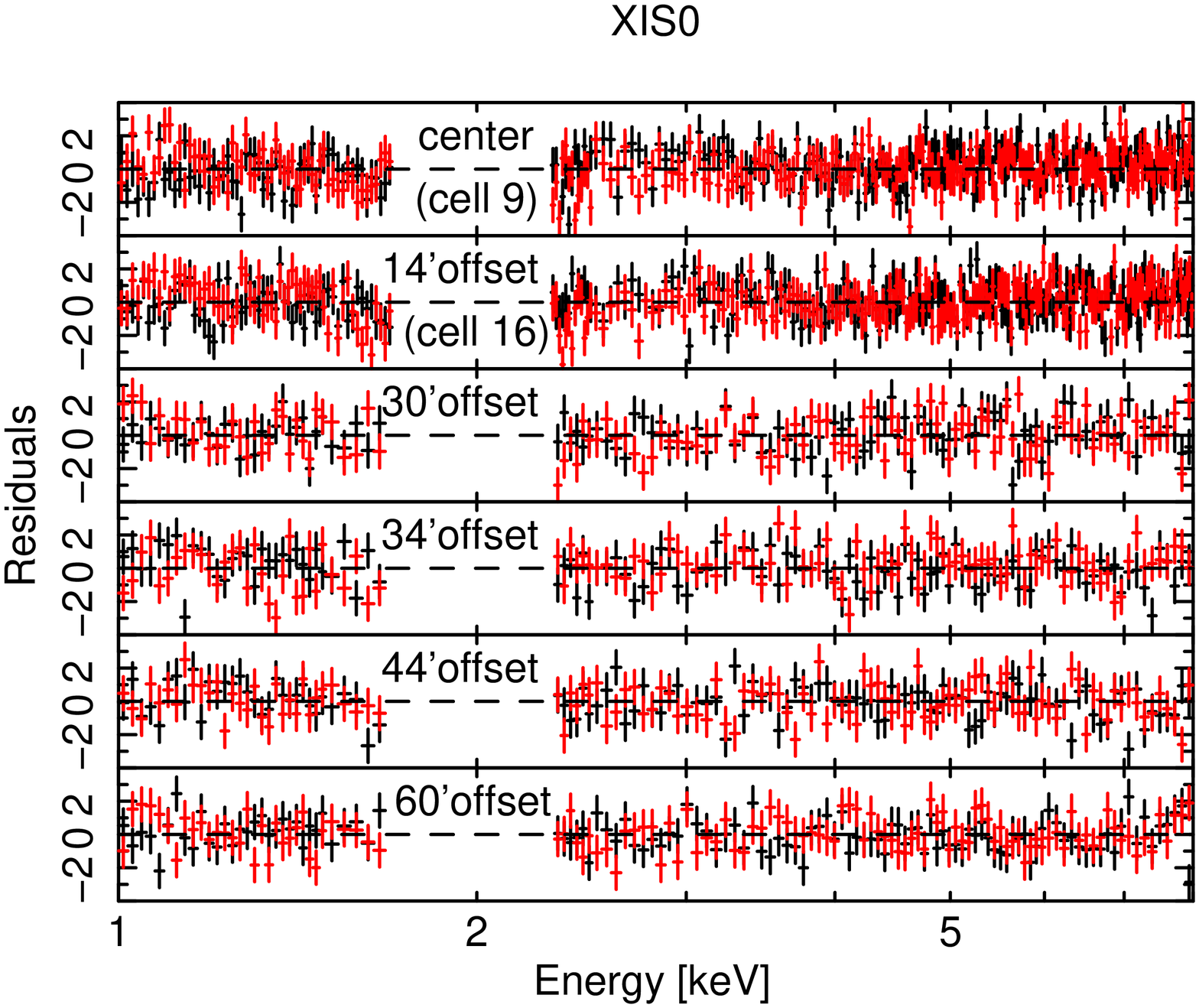}
\end{minipage}\hfill

\caption{Left: Comparison of $\chi^{2}$ fitted with 
single and three temperature (APEC) models. Color notations 
are the same as those in figure \ref{fig:image}. Right: Residuals of the fit
of XIS0 with single--temperature (black) and three-temperature (red) models
versus energy. Each panel shows the region of the spectral fit.
}
\label{fig:3Tchi}
\end{figure*}

As shown in figure \ref{fig:spec}, the spectra in the
energy range of 1.0--8.0 keV are well-represented 
by the single-temperature model. 
Except for discrepancies between FI and BI detectors around the Si-edge, 
there are no systematic residuals. 
Figure \ref{fig:3Tchi} shows a comparison of the resultant 
$\chi^2$ fitted with single and three temperature (APEC) models
in the energy range of 1.0--8.0 keV, excluding the 1.7--2.3 keV energy range.
A more statistically rigorous approach would be to apply the f-test
to compare single-temperature and three temperature models. The f-test indicates that the
three-temperature model is significantly better than the single-temperature model in regions 
1, 5, 6, 7, 9, 10, and 11 of the center region and 17 of the 14$'$ offset region,
 and better in regions 2, and 14 of the center region, and 16, and 19 of the 14$'$ offset region,
 30$'$ offset region, 44$'$ offset region, and 60$'$ offset region. 
The largest f statistic value is 11.4 in cell number 7 of the center region,
and the least f statistic value is 0.238 in cell number 8 of the center region.
However, the single-temperature model fit still represents the spectra fairly well.

\subsection{Temperatures derived from Fe line ratios}

\begin{figure*}[th]

\begin{minipage}{0.45\textwidth}
\FigureFile(80mm,60mm){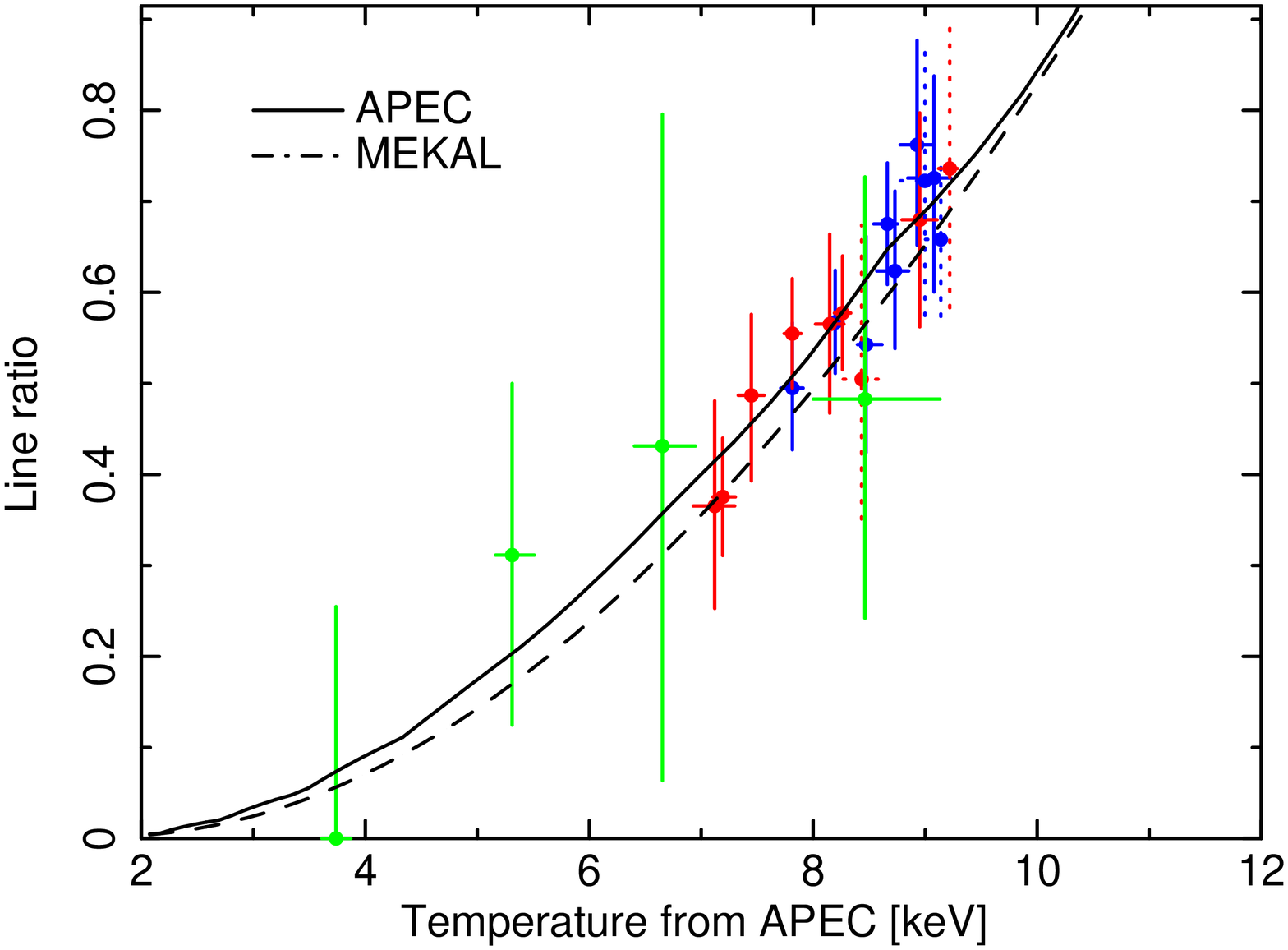}
\end{minipage}\hfill
\begin{minipage}{0.45\textwidth}
\FigureFile(80mm,60mm){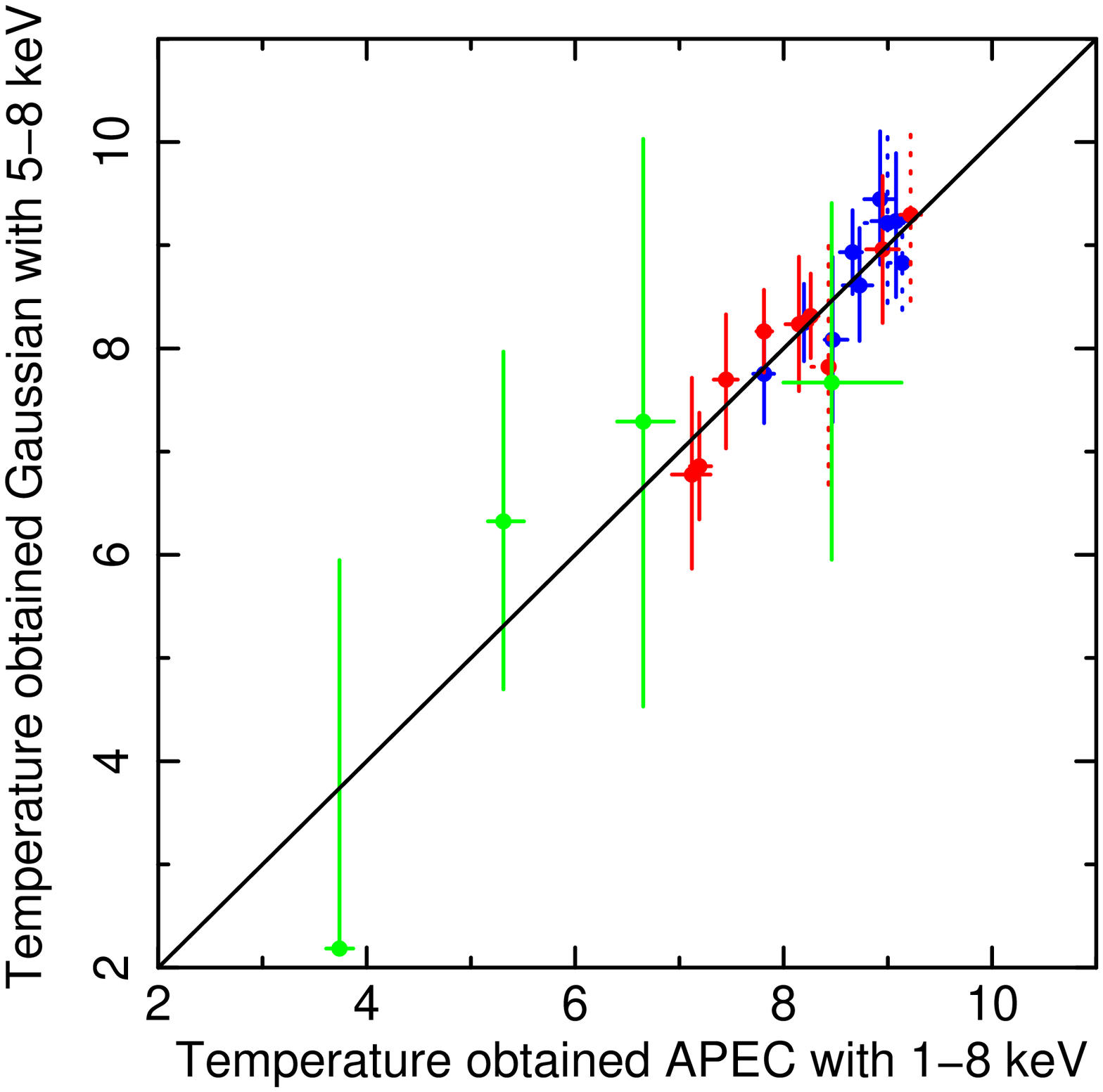}
\end{minipage}\hfill

\caption{
Left: Comparison between ratios of the normalizations 
of Fe K$\alpha$ lines of H-like and He-like Fe and the
temperatures derived from the spectral fits with the single 
temperature (APEC) model in the energy range of 1.0-8.0 keV\@. Closed blue and red circles 
with solid error bars correspond to the results of the north and 
south cells in the center region, respectively. Those with dashed 
error bars show the results of northeast and southwest cells 
in the 14$'$ offset region, respectively. Green circles indicate 
the results of the four offset regions. Solid and dashed black 
curves show the theoretical relation between the line ratio and 
plasma temperature from APEC and MEKAL codes, respectively. 
Right: Comparison of temperatures derived from line ratios
of Fe K$\alpha$ lines and spectral fits with the 
single temperature (APEC) model in the energy range of 1.0--8.0 keV\@.
}
\label{fig:kT_ratio}
\end{figure*}

The spectra accumulated over each region may contain multi 
temperature components, although the single temperature (APEC) 
model represented those spectra well. Because the ratio of 
He-like and H-like Fe K$\alpha$ lines strongly depends on 
the plasma temperature, comparisons of temperatures derived 
from the line ratio and spectral fits with the thermal model 
are useful for understanding the temperature structure of ICM.
Figure \ref{fig:kT_ratio} shows the ratios of the resultant 
normalizations of the Fe K$\alpha$ lines, plotted versus the temperature 
derived from the spectral fits with the single temperature (APEC) model 
in the energy range of 1.0--8.0 keV\@. To convert line ratios to plasma 
temperature, we generated mock spectra assuming the single temperature 
APEC or MEKAL \citep{Mewe1985,Mewe1986,Kaastra1992,
Liedahl1995} plasma code models, convolved with XIS energy resolution. 
We then fitted the mock spectra with a sum of a bremsstrahlung and three 
Gaussian models. The theoretical line ratios from APEC and MEKAL 
plasma codes are also plotted in figure \ref{fig:kT_ratio}.
The derived line ratios agreed with these theoretical 
line ratios of the APEC plasma code assuming the single temperature 
model, while the MEKAL model gave a line ratio value several \% lower 
at a given temperature. Assuming a single temperature 
plasma, we converted the observed line ratios to plasma temperatures 
using the theoretical relation for the APEC model. 
Results are shown in table \ref{tab:1-8vapecresults} and figure \ref{fig:kT_ratio}. 
As shown in figure \ref{fig:kT_ratio}, 
line temperatures agreed with those derived from the spectral fits very well.

\subsection{Temperatures at the edge of the radio halo }

\begin{figure*}[ht]
\begin{minipage}{0.45\textwidth}
\FigureFile(80mm, 60mm){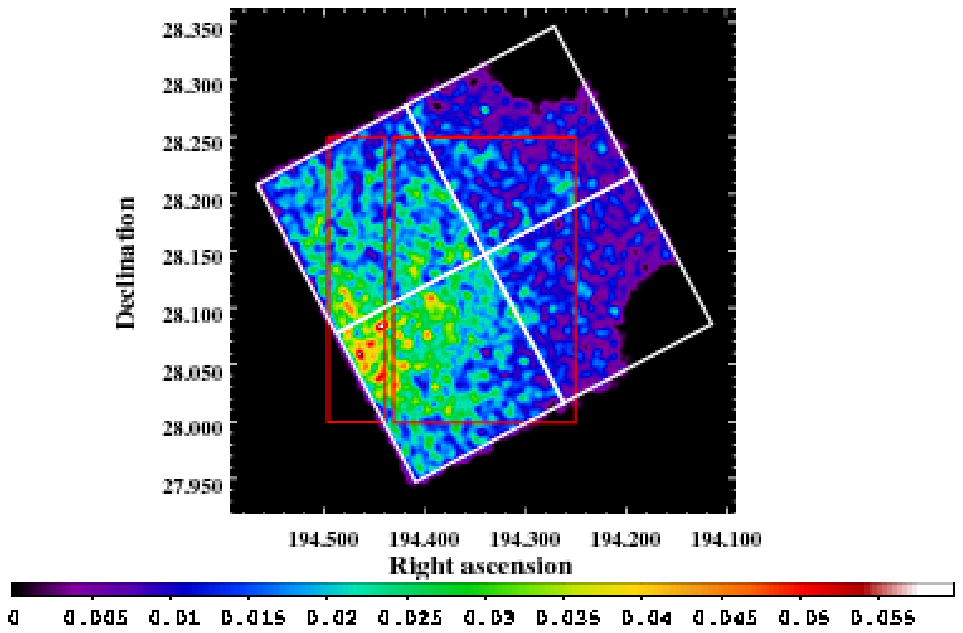}
\end{minipage}\hfill
\begin{minipage}{0.45\textwidth}
\FigureFile(80mm, 60mm){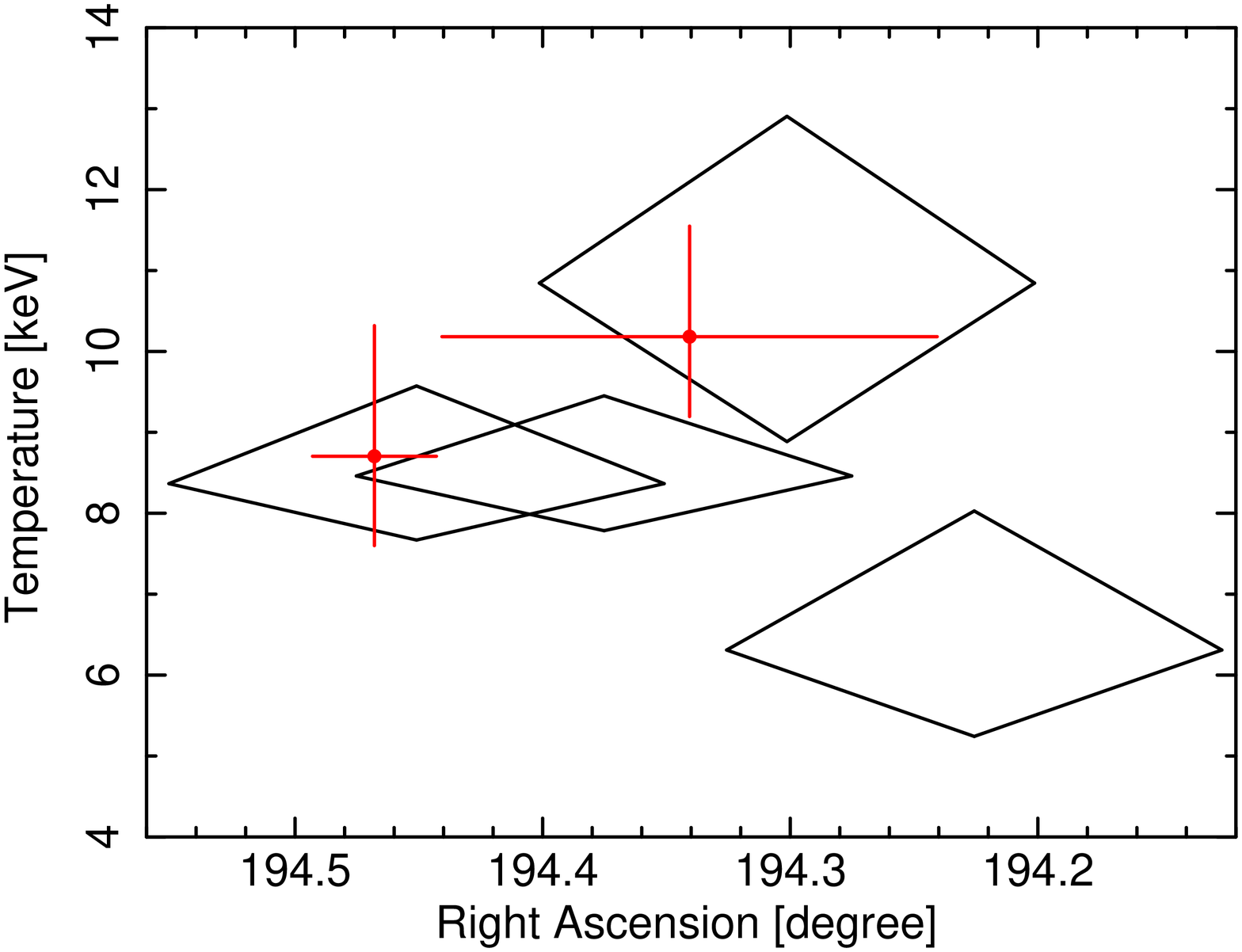}
\end{minipage}\hfill

\caption{
Left: XIS0 image of the 34$'$ offset region.
Red and white squares indicate spectral extraction regions.
Right: Temperatures of ICM in the 34$'$ offset region, plotted 
versus Right Ascension. Red crosses and black diamonds correspond 
to the red and white squares, respectively.
}
\label{fig:Fino}
\end{figure*}

\citet{brown10} claimed a significant temperature difference of ICM 
at the edge of the radio halo on the basis of XMM-Newton observations:
The temperature at the inner edge of the radio halo is 6.8$\pm$0.6 keV,
while the temperature at the outer edge of the radio halo is 16.3$\pm$2.9 keV.
To examine this temperature difference, 
we extracted spectra from counterpart regions in the 34$'$ offset region, 
which are represented by two red boxes in figure \ref{fig:Fino}, and fitted 
the spectra with the single temperature (APEC) model.
As shown in figure \ref{fig:Fino}, there are no significant temperature 
differences for each box in the 34$'$ offset region. We then divided FOV 
of the 34$'$ offset region into 2$\times2$ cells as shown in figure 
\ref{fig:Fino}, and fitted the spectra with the single temperature 
(APEC) model. The derived temperatures are also consistent with each other.

\subsection{Search for ICM Bulk Motions}
\label{subsec:bulk}

\subsubsection{Systematic uncertainty in the energy scale}
\label{subsec:cal}

To constrain the line-of-sight gas motion using the He-like
Fe-K$\alpha$ line, precise calibration of the XIS energy scale is
crucial. Because the six data sets used in the present analysis have
three different observation epochs (May 2006, June 2009 and December 2009), 
it is necessary to check the energy scale for each observation. 
In addition, the energy gain may vary from place to place
on the same CCD chip owing to the charge transfer inefficiency. We thus
estimate the XIS calibration uncertainty in the following two ways:
(i) measurements of the calibration source energy and (ii) comparisons
of redshift values derived from four or three sets of XIS.

\begin{figure*}[th]
\begin{minipage}{0.45\textwidth}
\FigureFile(80mm,60mm){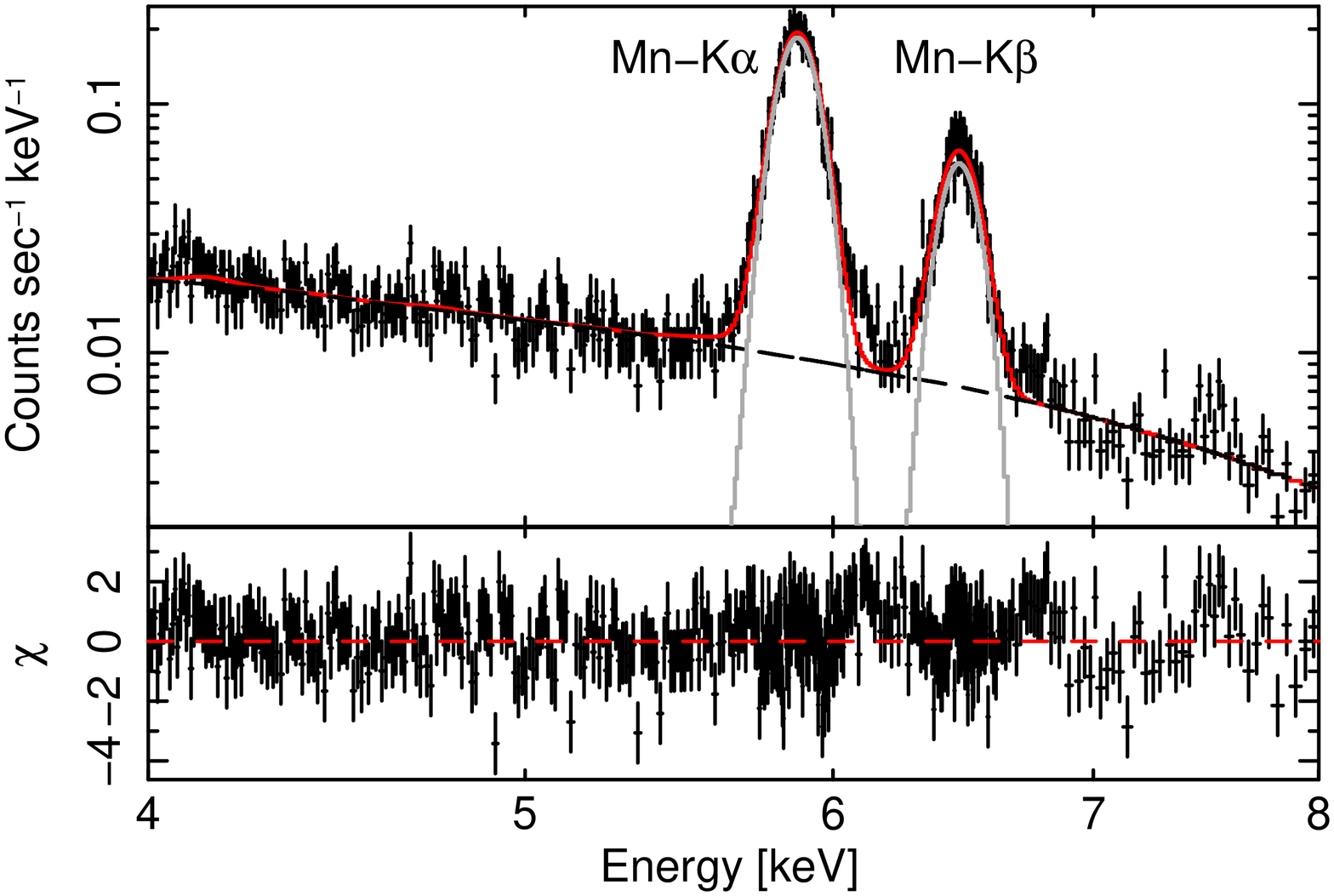}
\end{minipage}\hfill
\begin{minipage}{0.45\textwidth}
\FigureFile(80mm,60mm){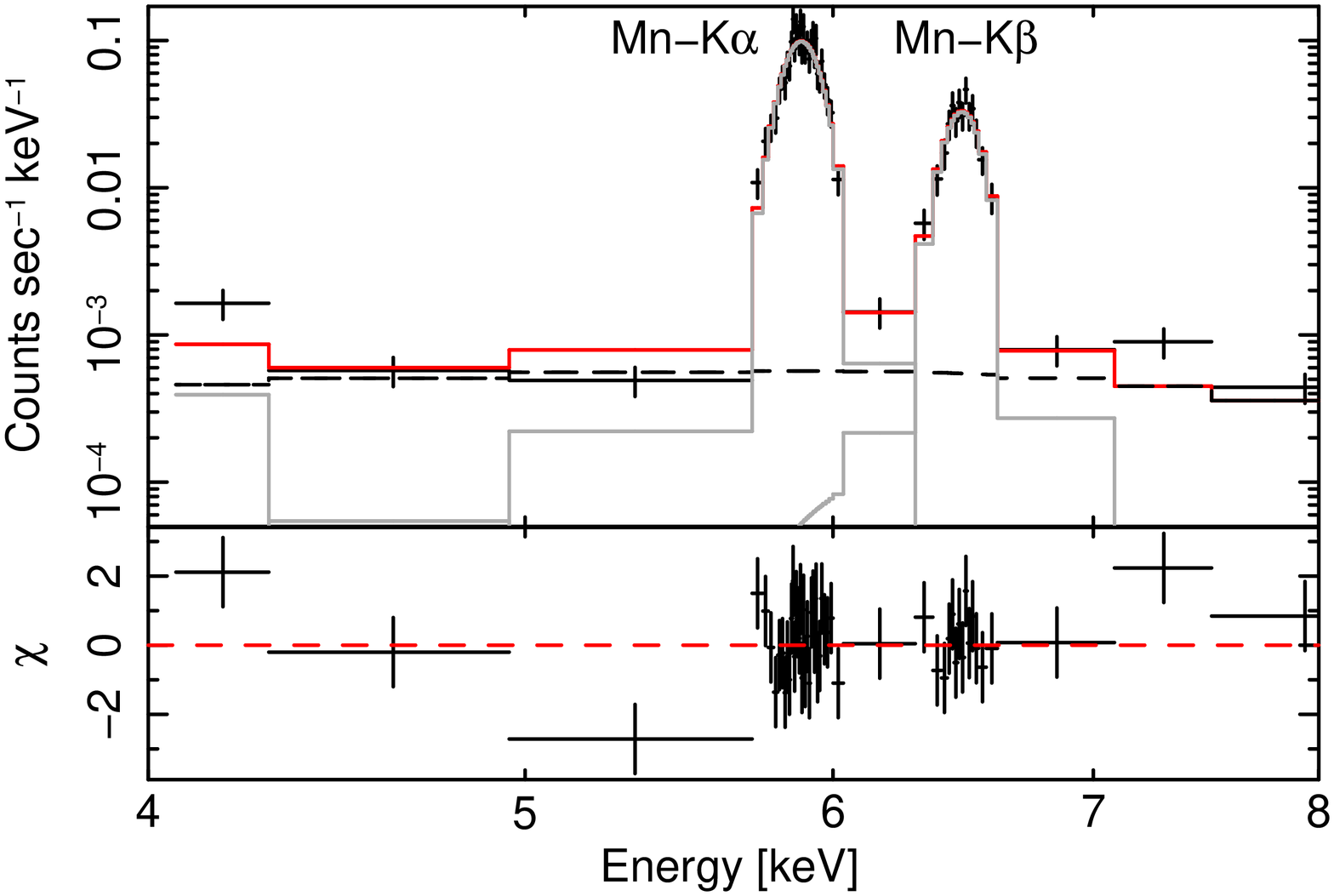}
\end{minipage}\hfill

\caption{Observed XIS0 spectrum (black crosses) in the energy range 
around the calibration source for the center (left) and 30$'$ 
offset (right) regions fitted with power-law and two Gaussian models.
The red line shows the best fit model and the dashed and solid gray lines 
indicate the bremsstrahlung and Gaussian models, respectively. 
Mn-K$\alpha$, Mn-K$\beta$ lines, and Fe lines from the cluster emission 
are seen in this spectrum. The lower panels show the residual of the fits.}
\label{fig:cal}

\begin{minipage}{0.45\textwidth}
\FigureFile(80mm,60mm){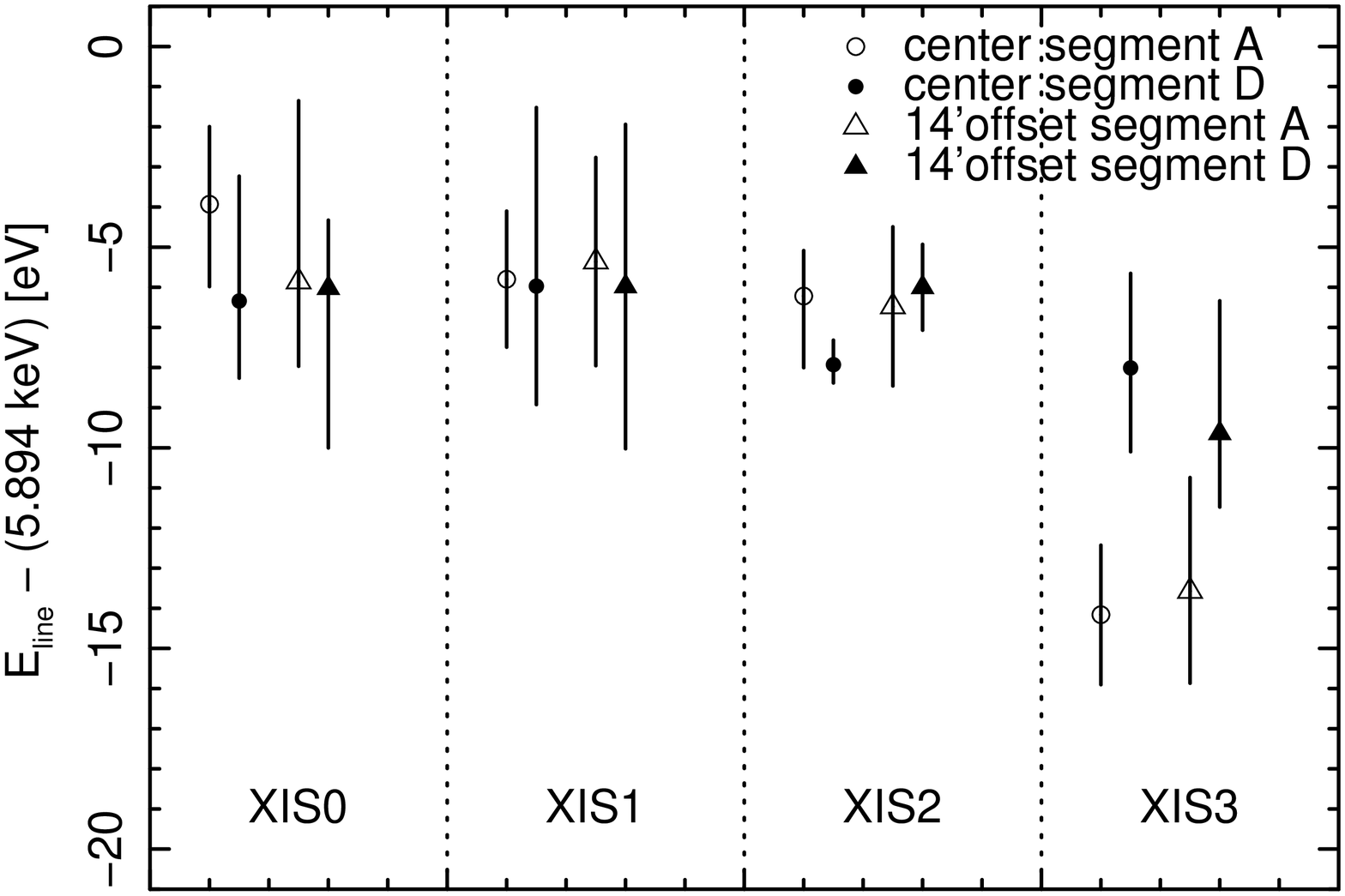}
\end{minipage}\hfill
\begin{minipage}{0.45\textwidth}
\FigureFile(80mm,60mm){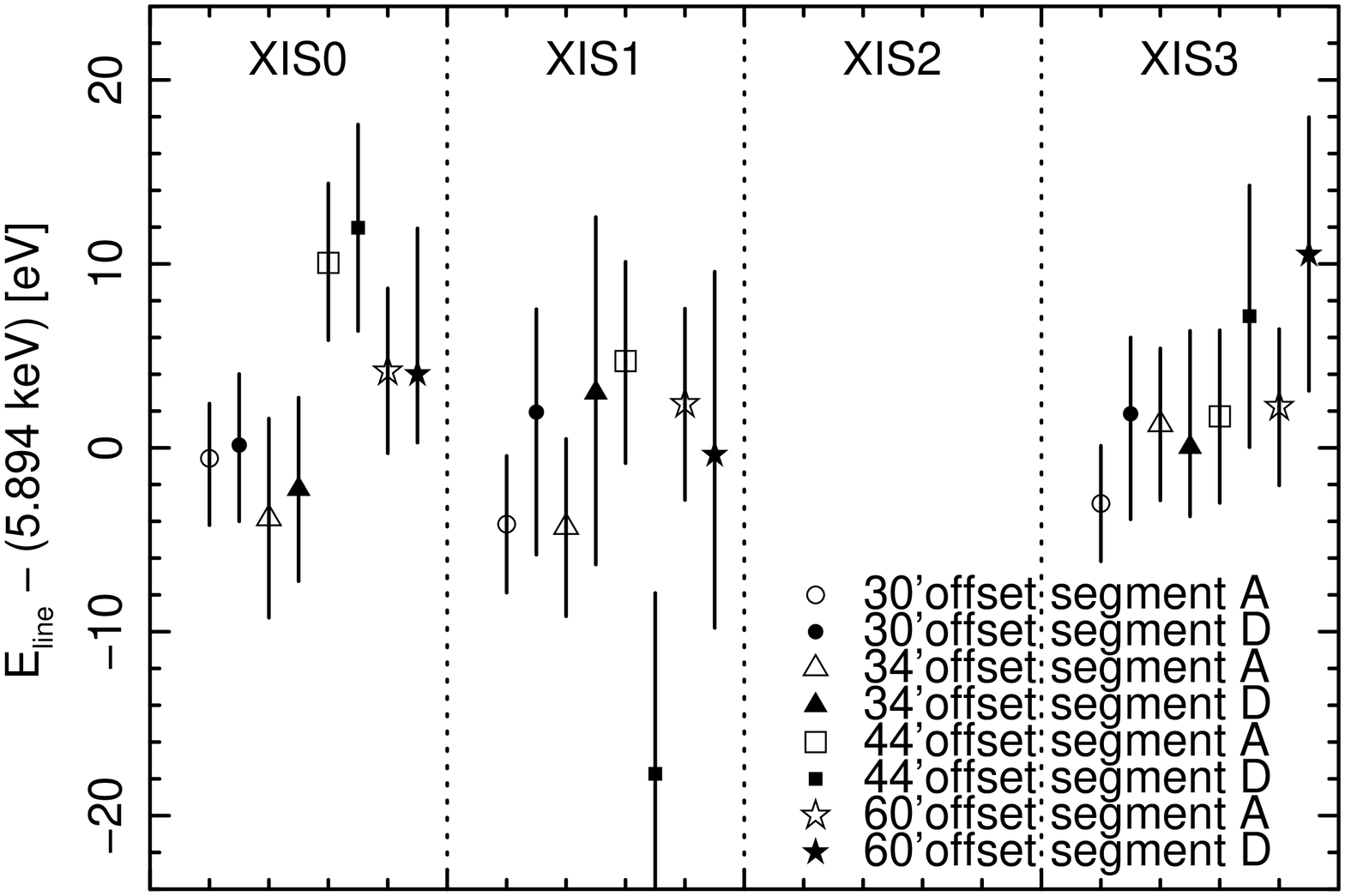}
\end{minipage}\hfill

\caption{Left: Resultant residuals of the central energy of 
the Mn-K$\alpha$ line (5.894 keV) for the center and 14$'$ offset 
regions. Clear and shaded symbols show the segments A and D on XIS, 
respectively. Right: Same as the left panel for the 30$'$, 34$'$, 
44$'$, and 60$'$ offset regions.
}
\label{fig:cal_result}
\end{figure*}

(i) The fiducial absolute energy scale can be examined by measuring
the centroid energy of the Mn K$\alpha$ line from built-in
calibration sources, which illuminate two corners of each XIS chip.
The spectra accumulated from the calibration source regions were
fitted to the power-law model for continuum and two Gaussian functions
for the Mn K$\alpha$ line at 5.894~keV and Mn K$\beta$ line at
6.490~keV. Because the Mn K$\beta$ line has an intrinsically complex line shape and is
contaminated by the cluster iron emission, only the Mn K$\alpha$ line 
is used for the calibration. Figures~\ref{fig:cal} and
\ref{fig:cal_result} show the result of spectral fitting and the 
Mn K$\alpha$ line centroid energy, respectively.

As shown in the left panel of figure~\ref{fig:cal_result}, measured line
energies for the center or 14$'$ offset regions are systematically
lower than the Mn K$\alpha$ line energy by 7.3~eV if they are averaged over four
sensors and two detector segments in each pointing. For XIS3,
centroid energies are lower than those of the other three chips,
however, we confirmed that the spectral analysis of the redshift measurement 
with and without XIS3 gave statistically consistent results 
(see section \ref{sec:resultred} for more details). 
We then estimate the $1\sigma$ systematic uncertainty on the
energy scale to be $\sigma_{\rm sys, Mn}=7$~eV. This is justified
also from the fact that if we calculate the $\chi^2$ value
using the measured Mn K$\alpha$ line centroid $E_i$ and the $1\sigma$ statistical error $\sigma_i$ 
in the XIS-i detector, and the Mn K$\alpha$ line energy in literature $\langle E \rangle$.
$\chi^2$ is defined as 
$\chi^2=\sum_{\rm seg}\sum_{i=0}^{3}(E_i - \langle$E$\rangle)^2/(\sigma_i^2 + \sigma_{\rm sys, Mn}^2)$.
With $\sigma_{\rm sys, Mn}=7$~eV, $\chi^2$ becomes less than 10 for 7 degrees of freedom, 
which means the probability of $\chi^2$ exceeding $\chi^2$ versus the number of 
degrees of freedom becomes less than 10\%: 
the number of bins is 8 for four sensors and two detector segments, 
and $E$ can be regarded as being consistent with $\langle E \rangle$ 
when we consider $\sigma_{\rm sys, Mn}=7$~eV.
In the same manner, the systematic errors are obtained 
as $\sigma_{\rm sys, Mn}=1$~eV, $1$~eV, $6$~eV, $4$~eV for 
$30'$, 34$'$, $44'$, and $60'$ offset regions, respectively. 
Therefore, for the safety's sake, we assign 7~eV for the 1$\sigma$ 
(68\% confidence level) gain uncertainty.

(ii) Though calibration sources provide information on the absolute
energy scale at the corners of CCD chips, the gain may be dependent on 
position. This intrachip variation can be effectively
studied by comparing line energies of the same sky regions (which
correspond to different detector regions) on four or three XIS
sensors (see also section~3.1 in \cite{ota07}).
To examine this issue, we divided the center region into 16 cells,
and fitted the $APEC$ model in the energy range of 5.0--8.0 keV for each XIS sensor.
We then estimate the 1$\sigma$ systematic error with a reference to 
obtained redshift, calculating as 
$\chi^2=\sum_{k=0}^{15}\sum_{i=0}^{3}(E_{z_k^{(XIS-i)}} - \langle E_{z_k}\rangle)^2
/({\sigma_k^{(XIS-i)}}^2 + \sigma_{\rm sys}^2)$
for each pair of XIS sensors, where $E_{z_k^{(XIS-i)}}$ is the measured line centroid 
for the XIS-i detector in each region, $\langle E_{z_k}\rangle$ is the mean of the line centroid 
for all XIS-i detectors in each region, and $\sigma_k^{(XIS-i)}$ 
is the $1\sigma$ statistical error in the XIS-i detector.
With $\sigma_{\rm sys}=$11 eV, $\chi^{2}$ becomes less than 17 
for 11 degrees of freedom, which means the probability of $\chi^{2}$ exceeding 
$\chi^{2}$ versus the number of degrees of freedom becomes less than 10\%:
the number of bins is 12 for the spectral regions except for the calibration source
region in the center region.
$\sigma_k^{(XIS-i)}$ is $\sim$ 10 eV, which is larger than the gain
uncertainty: thus, we ignored $\sigma_k^{(XIS-i)}$ to avoid underestimating the systematic error. 
Because $\sigma_k^{(XIS-i)}$ is ignored in these estimates, 
the obtained value is an upper limit for intrachip variation within a 68\% confidence level.

In summary, the intrachip variation is an upper limit and is larger than the gain uncertainty.
Thus, we concluded from (i) and (ii) that the 1$\sigma$ systematic error 
within a 68\% confidence level is 11 eV.
The systematic error is then 18 eV, which corresponds to 818~km~s$^{-1}$ in the line of sight velocity,
 when we quote a 90$\%$ confidence level.

\subsubsection{Redshift Measurements from Two Different Spectral Models}

\begin{figure*}[htbp]

\begin{minipage}{0.45\textwidth}
\FigureFile(80mm, 60mm){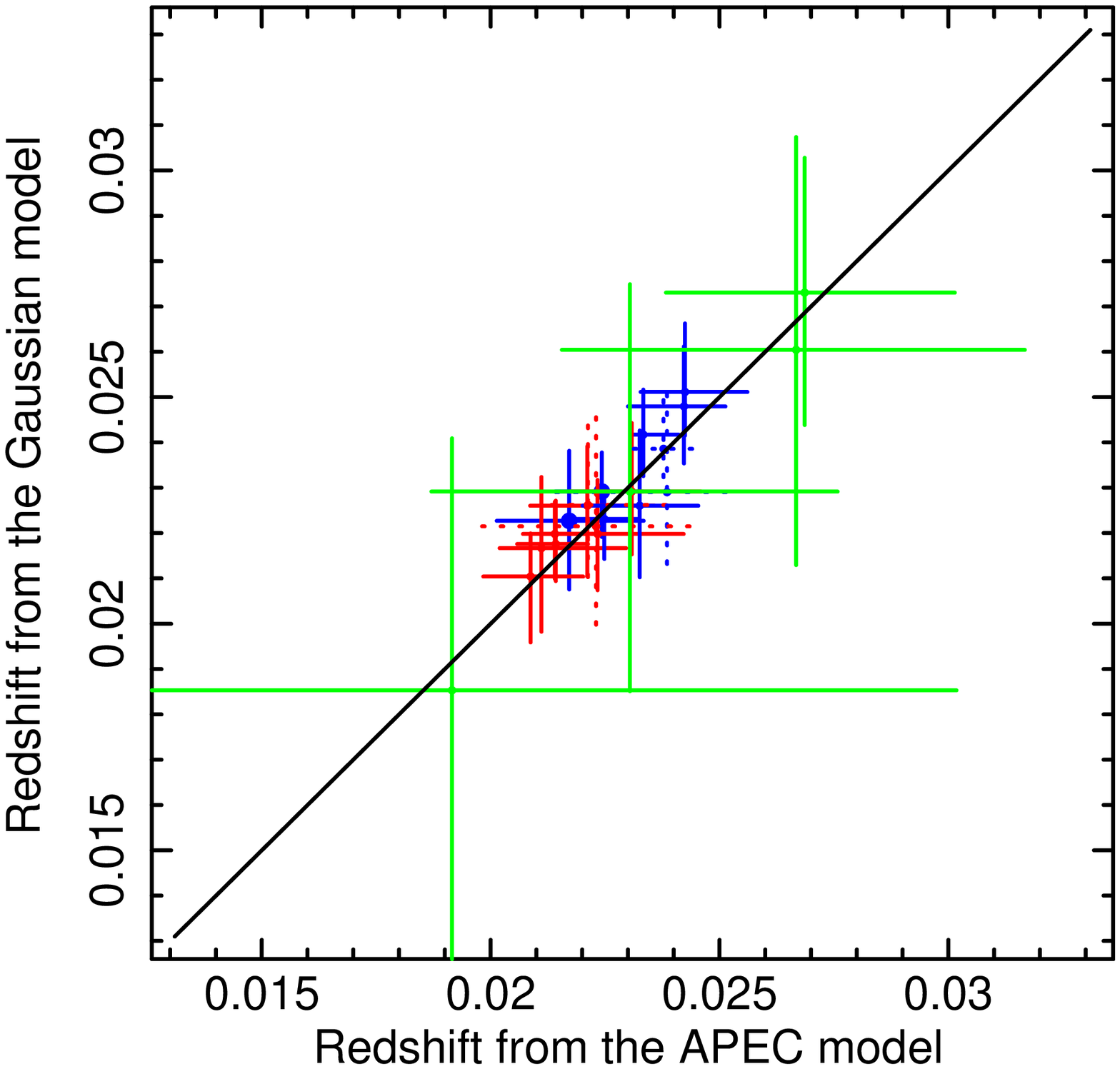}
\end{minipage}\hfill
\begin{minipage}{0.45\textwidth}
\FigureFile(80mm, 60mm){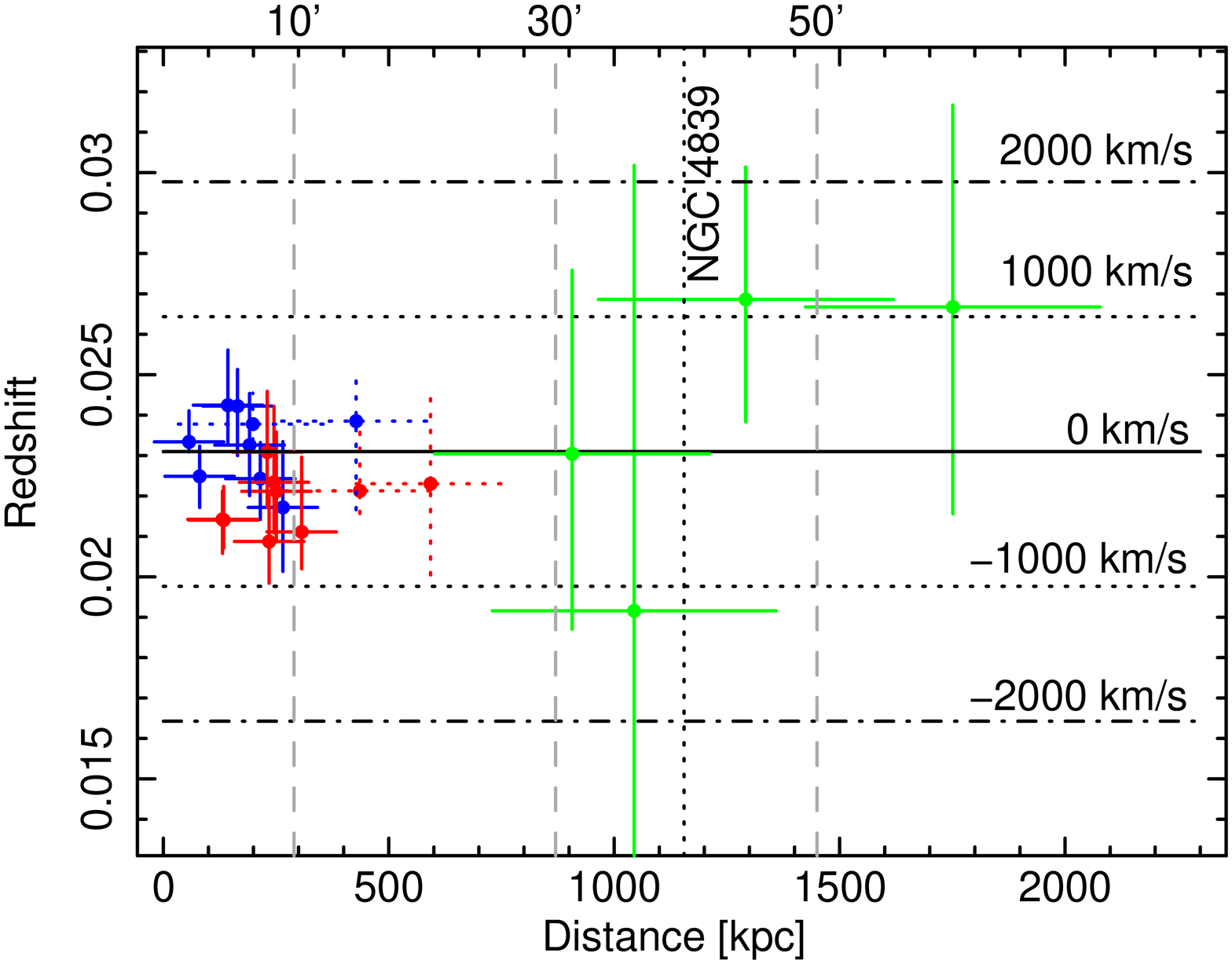}
\end{minipage}\hfill

\caption{
Left: Comparison of redshifts derived from spectral fits with 
the single temperature (APEC) model in the 5.0--8.0 keV energy range\@, 
and the shifted centroid energy of the He-like Fe K$\alpha$ line.
Right: Derived redshifts with the single temperature (APEC) model 
in the 5.0--8.0 keV energy range\@, plotted versus the distance from the X-ray peak 
of the Coma cluster. Color notations are the same as in figure 
\ref{fig:image}. The solid line corresponds to the optical redshift 
of the Coma cluster, and the dotted and dash-dotted lines show 
the redshift of $\pm$1000 and $\pm$2000 km, respectively.
}
\label{fig:redshift}
\end{figure*}

As mentioned in subsection \ref{subsec:single} and \ref{subsec:ratios}, 
the line-of-sight velocity of ICM or redshift can be 
evaluated using two different spectral models: the APEC model 
or the bremsstrahlung with three Gaussian lines. It is then useful 
to compare these results and see if there is any 
systematic uncertainty in the redshift measurement due to 
spectral modeling. In the APEC model, the temperature dependence 
of the Fe K$\alpha$ line energy is already implemented in the 
plasma code, which in turn evokes a parameter coupling between 
the temperature and line energy (or redshift). On the other hand, 
in the bremsstrahlung$+$Gaussian model, the line centroid should be
determined almost independently of the continuum temperature, although
its temperature dependence has to be corrected later. We did
this correction by deriving the rest-frame Fe K$\alpha$ centroid
energy as a function of temperature on the basis of spectral simulations,
assuming the APEC model and XIS energy responses.

In figure~\ref{fig:redshift}, the comparison of redshift measured 
by the two different models is shown for all analysis regions 
defined in figure~\ref{fig:image}. Because there consistency 
between the two within their statistical errors, we conclude that 
either method robustly determines the ICM redshift for 
the present observations. Therefore, we will show 
results of the fits with the APEC model because the uncertainties with APEC are smaller.

\subsubsection{Redshift Measurement from the APEC model}
\label{sec:resultred}

The obtained redshifts from the spectral fits with the single 
temperature (APEC) model in the energy range of 5.0--8.0 keV are shown in  
figure~\ref{fig:redshift}, where the ICM velocity in the line
of sight is also calculated as $v_{\rm l} \equiv c(z - z_{\rm cl})$\@.
The resultant redshifts are consistent with the optical redshift
in \citet{colless1996} from the NED data base\footnote{http://nedwww.ipac.caltech.edu/}, 
within statistical and systematic errors, which corresponds 
to $\pm$0.0027 in the redshift. The direct measurements of the 
central energy of the K$\alpha$ line of He-like Fe also give
almost the same redshifts derived from the spectral fits with the 
single temperature (APEC) model.
To quantitatively derive the redshift, we calculated 
the mean redshift of the ICM in all pointings to be
$\langle z_{obs} \rangle= 0.0227\pm0.0005$. This is slightly lower 
than the optical redshift, $z_{\rm cl}=0.0231$ from the NED database,
but within 90$\%$ error.
As shown in table \ref{tab:1-8vapecresults}, in the center and 14$'$ offset regions, 
the derived highest and lowest redshift are $z_{\rm highest}=0.0242\pm0.001$
 for cell numbers 5 and 8 of the center region, and $z_{\rm lowest}=0.0209\pm0.001$ 
for cell number 2 of the center region. 
Based on these redshift measurements, the derived ICM bulk velocities are 
$v_{\rm highest}=330\pm300$ km s$^{-1}$, and 
$v_{\rm lowest}=-660\pm300$ km s$^{-1}$.
Including systematic errors of $\pm818$ km s$^{-1}$ with 90$\%$ confidence level (Section \ref{subsec:cal}), 
there is no significant difference. We estimate the 90$\%$ upper limit on the 
velocity gradient to be $\Delta v=990\pm424\pm1157$ km s$^{-1}$\@.
If we further add systematic and statistical errors in the quadrature, the upper limit of the gas velocity
within the 90$\%$ confidence level becomes $|\Delta v|$ $<$ 2200~km~s$^{-1}$.
For simplicity, assuming that the gas is rigidly rotating at a typical circular velocity of 
$\sigma_{r} \sim |\Delta v|/2$, then $\sigma_{r}=1100$~km~s$^{-1}$, which does not exceed the sound velocity
of the Coma cluster, which is 1500~km~s$^{-1}$ for a temperature of 8~keV.
The average redshift in the center region is as follows:
$0.0227\pm0.0011$ in the north-west region (cell numbers 8, 9, 12, and 13), $0.0214\pm0.0013$ 
in the south-west region (cell numbers 10, 11, 14, and 15), 
$0.0236\pm0.0012$ in the north-east region (cell numbers 0, 1, 4, and 5), 
and $0.0222\pm0.0012$ in the south-east region (cell numbers 2, 3, 6, and 7).
We also show the averaged redshift over FOV in the upper panel in table \ref{tab:redshift}. 
These results are consistent with each other in the center region, 
and with the observed redshifts in the 14$'$ offset region 
within statistical and systematic errors.
The difference in the ICM velocity between the center 
and 30$'$ offset regions is derived as $\Delta v=90\pm1350\pm818$ km s$^{-1}$\@,
and the difference between the center and 34$'$ offset regions is $\Delta v=-1050\pm3300\pm818$ km s$^{-1}$\@.
These are also consistent with each other.
Components of the velocity in the plane of the sky are not measured. 
If we assume that three components are similar, then the limit on the radial velocity translates
into a limit of $\sqrt{3}\times 290 \sim$ 500 km~s$^{-1}$, which does not exceed the speed of sound
and is consistent with systematic errors.
Therefore, there were no significant velocity variations on the 10$'$ scale within central regions
(center and 14$'$, 30$'$, and 34$'$ offset regions) from our analysis. 
In the NGC~4839 subcluster region (44$'$ offset region), the observed redshift is 0.0269$\pm$0.0033, 
which corresponds to 1140$\pm990$ km s$^{-1}$ as the ICM bulk velocity without systematic errors. 
There also seems to be no significant difference between the core 
of the Coma cluster and the NGC 4839 subcluster within the 90\% confidence level.

As mentioned in subsection~\ref{subsec:cal}, some sensors have large uncertainties 
in the determination of the line centroid energy. On XIS3 in the center 
region, the Mn K$\alpha$ line centroid energy of segment~A was 
different from others. Similarly, the segment D of XIS1 in the NGC~4839 
region was also different. We averaged the spectrum over FOV
in (i) all detectors, and (ii) all detectors except those with different Mn K$\alpha$ 
line centroid energies to check the spectral shape and measured redshift 
in the center and NGC~4839 regions (44$'$ offset region). 
The results are shown in the lower section in table \ref{tab:redshift}.
For the center region, the derived redshifts were fairly consistent 
including XIS3 or not. On the other hand, for the NGC~4839 subcluster 
region, the resultant redshift changed slightly from 0.0268 to 0.0287, 
although this was also consistent within statistical and 
systematic errors. As shown in figure~\ref{fig:cal_result},
the peak of the He-like Fe-K$\alpha$ line of XIS1 is more shifted than 
that of the other two sensors.
The measured redshifts in the 44$'$ offset region were 0.0271$\pm$0.0023, 
0.0303$\pm$0.0051, and 0.0255$\pm$0.0027 for XIS0, 1, and 3, respectively.
Because the measured redshift of the averaged spectra between XIS0 and 3 
was consistent with the redshift from simultaneous spectral fits, 
the value 0.0269 would be plausible.

In conclusion, there was no significant difference in the ICM 
velocity between the core of the Coma cluster and the NGC~4839
subcluster within 90\% confidence.

\begin{table*}[th]
\caption{Results of the redshift measurement with the single 
temperature (APEC) model in the energy range of 5.0-8.0 keV\@. The upper section shows 
the averaged redshift for each region with all XIS sensors 
simultaneously. The lower section shows the redshift in each field
with the averaged spectral fits. XIS detectors that are included are shown in parentheses.
Systematic errors of $\pm$818~km~s$^{-1}$ in the X-ray ($cz$) were not included in this table.
Optical redshifts are referred in \citet{colless1996}.
}
\begin{center}
\begin{tabular}{ccccc} 
\hline
field & X-ray ($z$) & Optical ($z$) & X-ray ($cz$) & Optical ($cz$) \\
& & & km s$^{-1}$ & km s$^{-1}$ \\
\hline
all pointings & 0.0227$\pm$0.0005 & 0.02307 & 6810$\pm$150 & 6917 \\
center & 0.0227$\pm$0.0005 & 0.02286 & 6810$\pm$150 & 6853 \\
14$'$ offset &0.0230$\pm$0.001 & 0.02286 & 6900$\pm$300 & 6853 \\
center and 14$'$ offset & 0.0227$\pm$0.0005 & 0.02286 & 6810$\pm$150 & 6853 \\
NGC~4839	& 0.0269$\pm$0.0033 & 0.02448 & 8070$\pm$990 & 7339 \\
\hline
center (XIS0123) & 0.0226$\pm$0.0005 & 0.02286 & 6775$\pm$150 & 6853 \\
center (XIS012) & 0.0226$\pm$0.0005 & 0.02286 & 6775$\pm$150 & 6853 \\
NGC~4839 (XIS013)	& 0.0287$\pm$0.0033 & 0.02448 & 8604$\pm$990 & 7339 \\
NGC~4839 (XIS03)	& 0.0268$\pm$0.0033 & 0.02448 & 8034$\pm$990 & 7339 \\
\hline\\[-1ex]
\label{tab:redshift}
\end{tabular}
\end{center}
\end{table*}

\subsection{Fe abundance of ICM}
\label{subsec:iron}

\begin{figure*}[ht]

\begin{minipage}{0.45\textwidth}
\FigureFile(80mm, 60mm){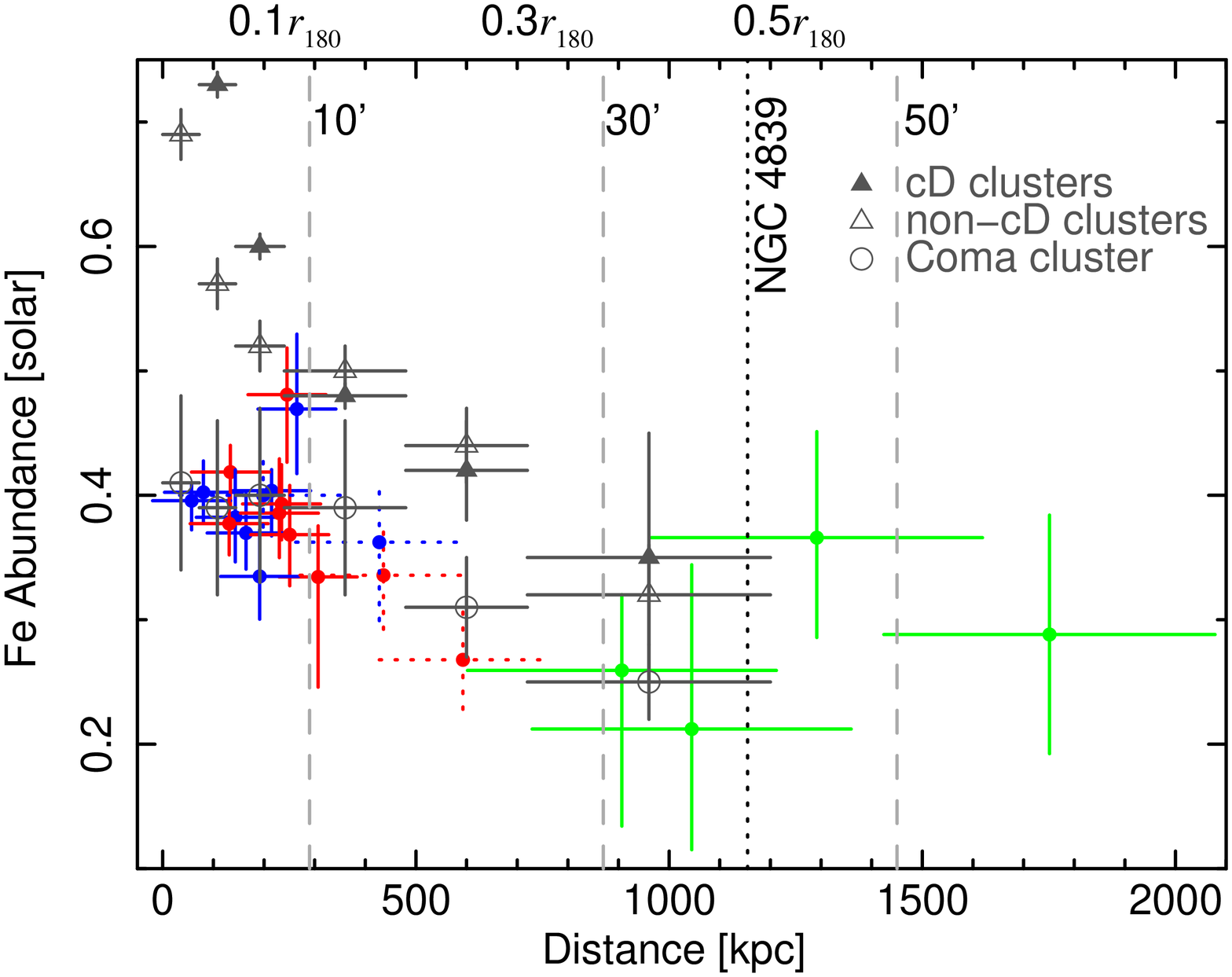}
\end{minipage}\hfill
\begin{minipage}{0.45\textwidth}
\FigureFile(80mm, 60mm){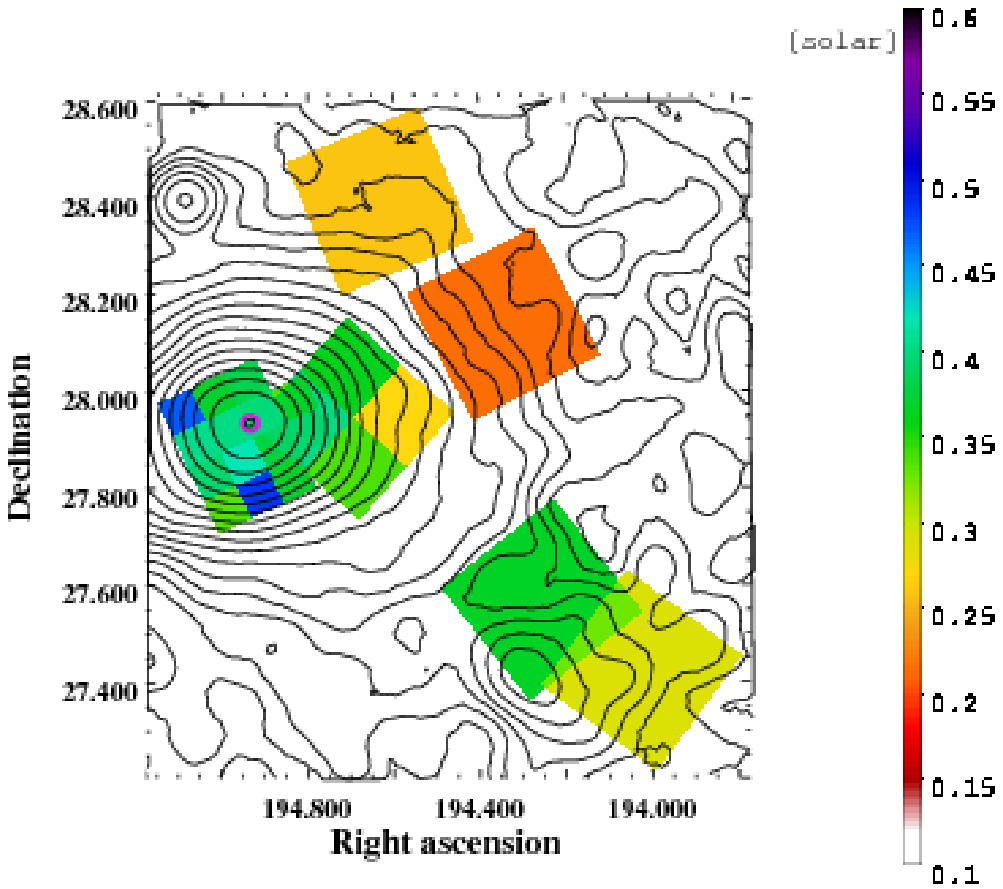}
\end{minipage}\hfill

\caption{Left: Radial Fe abundance profile with the single temperature 
(APEC) model in the 5.0--8.0 keV energy range measured from the position of the X-ray peak of the Coma cluster. 
Color and symbol notations are the same as in figure~\ref{fig:kT_apec}.
The gray colors show the results from XMM-Newton \citep{Matsushita2011}.
Right: Observed Fe abundance map of the Coma cluster in solar units. 
The magenta circle corresponds to the X-ray peak. The Fe abundance map 
is overlaid on X-ray contours on a linear scale in the 0.4--2.4 keV energy range 
with the ROSAT-All-Sky-Survey. }
\label{fig:Fe}
\end{figure*}

To measure the abundance using the strong Fe--K lines, 
we employed the results of spectral fitting in the 5.0--8.0 keV
energy range for the Fe abundance. 
Because of the good statistics, the spectra in the lower energy band primarily determine
the fitted temperatures. Although the spectra are well fitted with a single-temperature 
model, a small discrepancy between the model and data around the Fe-K line
can yield a large systematic uncertainty in the derived Fe abundances.
Within the 34$'$ region, temperatures and abundances derived from the spectral fitting
in the 5.0--8.0 keV energy range agree well with those from 1.0--8.0 keV
energy ranges within 5\%.
In the 44$'$ and 60$'$ offset regions, the Fe abundances from the 5.0--8.0 keV
energy range are smaller than those from the 1.0--8.0 keV range by $\sim$0.1 solar.
$\chi^{2}$ values for the limited energy range of 5.0--8.0 keV from the former
fitting are 10\% smaller than those from the latter, although the 
single-temperature model reproduces the observed spectra at 1.0--8.0 keV well.
Therefore, we adopted the Fe abundance derived from the spectral fitting
in the 5.0--8.0 keV energy range. 
Figure~\ref{fig:Fe} shows a radial profile and map of the Fe 
abundance of ICM with the single temperature (APEC) model 
in the 5.0--8.0 keV energy range\@. Within 30$'$ from the X-ray peak, 
Fe abundances are almost constant at $\sim$0.4 solar.
The derived Fe abundances are 0.41$\pm$0.01 solar 
in the center region, and 0.38$\pm$0.02 solar in the 14$'$ offset 
region. The Fe abundances in the 30$'$ and 34$'$ offset regions 
are slightly lower, $\sim$0.3 solar, although it has larger uncertainties. 
These values agree with recent XMM results 
\citep{arnaud01,Matsushita2011} within 1.2 Mpc.

The Fe abundance in the NGC~4839 subcluster region is comparable 
to that in the central region, and higher than that in the 30$'$ 
and 34$'$ offset regions.
The average value in the subcluster region is 0.33$\pm$0.09 solar.
This value is also consistent with that in \citet{neumann2001}, 
although the error bars in their results are much larger than those in our results.

\section{Discussion}
\label{sec:discuss}

\subsection{Temperature measurements with Suzaku}

X-ray observations of
clusters of galaxies can play a major role in determining cosmological parameters,
because of the importance of measuring gravitational mass.
The ratio of He-like K$\alpha$ to H-like K$\alpha$ lines of Fe is a very
steep function of temperature, and has a different temperature dependence 
from that of the continuum. 
The continuum emission is affected 
by the systematic uncertainties in the response and
non-thermal emission, while the line ratio does not.
Thus, by comparing the fitted temperatures between these two methods, 
the effects of the systematic uncertainty in the response 
and/or of the non-thermal emission is increased.
Using the APEC plasma code,
the temperatures derived from the spectral fitting of the 1.0--8.0 keV energy range
and the line ratio of Fe agree within a few percent 
when we averaged all regions.
The systematic uncertainties in temperatures from the line ratio 
are in the level of the systematic difference between those from APEC and MEKAL,
which is about 3\%.
The consistency of observed temperatures from the continuum and 
those from line ratios supports the
accuracy of Suzaku measurements of plasma temperature.
In contrast, systematic differences of $\sim$ 10\% 
in the cluster temperature among Chandra, PN, and MOS were 
reported in relatively high--temperature clusters (\cite{Snowden2008}).
Even using the same instrument, the derived temperatures changed systematically by $\sim 10$\%
between calibrations \citep{Reese2010, Matsushita2011}.

A systematic uncertainty of several percent in response matrices of a detector
can also cause systematic uncertainties in the temperature structure.
\citet{Matsushita2011} found that
the temperature structures of ICM in the Coma
cluster derived from  multi-temperature fits with PN and MOS spectra were different,
although the two detectors gave consistent temperatures within several percent.
When they allowed higher temperature components over 10 keV in hot clusters including
the Coma cluster, using MOS data, Fe abundances from the He-like line increased
by several tens of percent, owing to artificial detection of a hot component,
while those derived by PN were unchanged.
However, with Suzaku, we found that the single-temperature model fits the 
spectra of the Coma cluster very well, and we do not need temperature components
above 10 keV or below 2 keV to fit the spectra.
The multi-temperature components observed by XMM were not needed.

In the outer regions of clusters at low surface brightness, 
uncertainties in the background of Chandra and XMM
also lead to systematic uncertainties in ICM temperatures.
Some analyses found that ICM temperatures decreased outward by half at 0.6$r_{180}$
(e.g., \cite{vikhlinin2005}; \cite{pratt2007}), while others found flatter
profiles (e.g., \cite{arnaud2005}).
Systematic uncertainties in the derived temperatures due to uncertainties
in the background with Suzaku observations are much smaller,
because Suzaku XIS has a much lower and more stable background.
Also, previous XMM and Chandra results of temperatures 
in the outer regions have statistical errors
more than 20\%, while those of Suzaku have errors less than 10\%.

\subsection{Effect on the search of a hard component}

We need precise measurements of ICM temperatures to search for a
non-thermal hard component as an excess of the thermal component.
On the basis of the ICM temperature map derived with XMM-PN, 
\citet{wik09} found that the Suzaku HXD-PIN spectrum was described with the 
thermal emission from ICM, and there was no significant evidence of excess hard emission.
The temperatures of the Coma cluster derived from Suzaku-XIS are
 consistent within error bars with those from the PN detector:
the weighted average of temperatures with statistical errors 
within the center field is close to
that derived with XMM-PN within a few percent. The temperatures of offset fields
are consistent with those from XMM-PN within error bars.
Thus, our results support the non-detection of non-thermal 
hard emissions by \citet{wik09}

\citet{ota2008} reported the existence of hot gas at $25.3^{+6.1}_{-4.5}$ keV in RX J1347.5-1145.
They concluded that the gas properties can be explained by a fairly recent (within the last 0.5 Gyr)
collision of two massive (5$\times$10$^{14}$ M$_{\odot}$) clusters with
bullet-like high velocity ($\Delta v \sim 4500$ km s$^{-1}$). 
In contrast, \citet{wik09} found that 
there was no extremely high temperature gas in the Coma cluster
 in the HXD-PIN spectrum up to 70 keV.
On the basis of the consistency of derived temperatures from the continuum and the line ratio
and the results from the spectral analysis of HXD-PIN,
we conclude  that the amount of very high temperature gas is rather small.
From the spatial variation of the derived temperatures, 
 there is no region with extremely hot temperature:
the derived temperatures range from 7.0 to 9.0 keV within the center,
14$'$ offset, and $34'$ offset regions.

The Coma cluster has a cluster-wide synchrotron radio halo.
If the electrons in the cluster were accelerated by a very strong shock that
 occurred owing to a recent merger, we might expect the cluster to contain
 more very hot gas, although it is difficult to make this argument
 quantitative. In any case, in the Coma cluster, relativistic
 electrons producing the radio halo were apparently generated without
 producing a large amount of very hot gas. An example of a process
 which might do this is the turbulent re-acceleration model (e.g., \cite{brunetti2001}).

\subsection{Temperature structure of the Coma cluster}

Within the center and \timeform{14'} offset regions the spatial variation of derived temperatures 
is relatively small, from 7 keV to 9 keV.
There are no extremely hot and cool regions.
We did not detect any temperature jump corresponding to a shock in the offset regions.
Within each region for the spectral analysis, the single-temperature model fit the spectrum.
Therefore, in the center of the Coma cluster, ICM is in a nearly relaxed state.

Although the variation was small, there was some asymmetric temperature structure,
like a hot region northwest of the center.
If the cluster had been an isolated system collapsing from self-gravity,
the temperature structures would be symmetric.
The observed asymmetry in the temperature distribution indicates that 
external effects or interactions, such as merging, have occurred.
Numerical simulations of the evolution of the cluster 
(e.g., \cite{roettiger1996}; \cite{norman1999}) 
show that nonaxisymmetric structures in the temperature distribution 
are erased several Gyr after each merger.
The presence of asymmetric temperature structures suggest that the central region
of the Coma cluster has experienced a merger, and 
the lack of hard component shows that its core have relaxed in the last several Gyr.

The observed cool region in the southeast of the center region 
coincides with the filamentary structure originating near 
NGC 4911 and NGC 4921 \citep{vikhlinin1997, donnelly1999}.
 \citet{arnaud01} reported that the temperature of this cold area ranges from 5 to 7 keV. 
This region mostly contain the stripped subcluster gas,
which is thought to be relatively cool in the initial state.
The numerical simulations (e.g., \cite{schindler1993}) suggest that 
the gas would maintain its temperature for several Gyr.
\citet{ishizaka1996} showed that even a single subcluster collision 
with supersonic velocity creates both hot and cool regions
in the main cluster. Although heating due to collisions occurs, 
cooling would also occur through the adiabatic expansion of the stripped gas.

\subsection{Searching for gas bulk motions}

The measurements of bulk motions in ICM are also important to derive
the gravitational mass of clusters of galaxies.
We constrained the upper limit of ICM bulk motions in the center 
and \timeform{14'} offset regions on the scale of 117 kpc (\timeform{4.5'}) and 234 kpc (9$'$), 
respectively, which correspond to the scales of regions for spectral accumulation
(see section \ref{sec:obs} and subsection \ref{sec:resultred}).
The upper limit of gas bulk motions is $|\Delta v|$ $<$ 1100 km s$^{-1}$. 
As summarized in 
table \ref{tab:redshift}, 
considering the systematic error of  18 eV, or 818 km s$^{-1}$ with a 90$\%$ confidence level (Section \ref{subsec:cal}), 
there are no significant bulk motions in the center and \timeform{14'} offset fields.
Compared with the sound velocity of the Coma cluster, 
$s=(\gamma kT/ \mu m_{p})^{1/2} \sim 1500$ km s$^{-1}$ for a temperature of $kT$ = 8 keV, 
the bulk-motion in the central region of the Coma cluster does not exceed the sound velocity.

Our results do not conflict with the discovery of pressure fluctuations in 
the center of the Coma cluster by \citet{schuecker04} which 
indicates that at least 10\% of the ICM pressure is in turbulent form.
We have not constrained the turbulence, and
the scale of the fluctuation ranges between 29 kpc and 64 kpc,
which is smaller than our scale of spectral analysis.

The velocity difference between the main cluster and the NGC 4839 subcluster 
from optical observations is 486 km s$^{-1}$.
Although the best-fit line-of-sight velocity of the subcluster is higher than 
that of the main cluster by $\sim$ 1000 km s$^{-1}$, the difference is comparable to statistical and
systematic errors.
Because the temperature of the region is about 5 keV,
the sound velocity is $\sim$ 1150 km s$^{-1}$.
Considering the error bars, we conclude that the relative line-of-sight gas velocity 
in the NGC 4839 subgroup region does not greatly exceed the sound velocity.

Numerical simulations indicate that a major merger between clusters raises the
temperature of ICM and induces a bulk velocity of 
 the order of 1000 km s$^{-1}$ (e.g., \cite{roettiger1996}; \cite{norman1999}).
In the late phase of mergers, turbulent motions develop in a simulation by \citet{takizawa05}.
The non-detection of bulk-motions in ICM of the Coma cluster
also indicates that the central region of the cluster is in a somewhat relaxed state.
If so, the non-thermal electrons relevant to the radio halo of the Coma cluster 
may have been accelerated by the intracluster turbulence rather than shocks.
In conclusion, the line-of-sight gas velocity of the Coma cluster in the observed region 
does not greatly exceed the sound velocity, and the central region of
the Coma cluster is in a nearly relaxed state.
This means that the assumption of hydrostatic equilibrium  is approximately valid in calculating the cluster mass of 
the Coma cluster.

\subsection{Fe Abundance map}

The Fe abundance is nearly constant at $\sim$ 0.4 solar within a clustocentric radius of 0.5 Mpc, 
and beyond that distance it decreases with the radius to 0.2--0.3 solar. 
This profile is consistent  with those
measured by XMM (\cite{arnaud01}; \cite{Matsushita2011}).
Suzaku has a lower level of background, and a slightly better
energy resolution: therefore, systematic uncertainties 
in Fe abundance are considered to be smaller.

Differences in abundance profiles of ICM in central regions of 
 clusters with and without cool cores have been reported (\cite{fukazawa2000},
\cite{degrandi2001}, \cite{Matsushita2011}).
\citet{Matsushita2011} showed an abundance difference between the two type of clusters
within 0.1$r_{180}$, while at 0.1--0.3$r_{180}$, the average values of
Fe abundances are both $\sim$ 0.4--0.5 solar, and are consistent with each other.

The Coma cluster is thought to have experienced a major merger in the recent past
and subsequently, the central region came to a somewhat relaxed state.
During a cluster merger, the mixing of ICM could destroy any central Fe peak.
In the first stage of merging, the Fe peak and cool core remain intact 
as observed in some ongoing merging clusters  (\cite{molendi2000}; \cite{sun2002}).
For example, Abell~2256, an ongoing merging cluster, 
 has an abundance gradient in the direction opposite of the 
merging direction owing to the remixing of the gas.
Numerical simulation also support this scenario.
 \citet{schindler2005} performed hydrodynamical simulations
with two different merger types of clusters: Cluster 1 has only small merger events, while Cluster 2 
undergoes a major merger (mass ratio 1:3). In Cluster 1, the abundance is more than $\sim$ 0.4 solar 
within a radius of 0.1 Mpc, and decreases from $\sim$ 0.3 solar to $\sim$ 0.1 solar with a decrease in radius.
The situation is different in Cluster 2.
In the Coma cluster, the Fe abundance is $\sim$ 0.4 solar and homogeneous within a radius of 0.5 Mpc. 
This value of the Fe abundance is similar to those at 0.2--0.3$r_{180}$ 
in other clusters as shown in figure~\ref{fig:Fe}.
Therefore, tn the Coma cluster, the merger have destroyed the cool core,
and mixed the ICM completely at least within 0.5 Mpc.

\section{Summary \& Conclusion}
\label{sec:sum}

The Coma cluster which was observed with the X-ray Imaging Spectrometer (XIS) onboard
Suzaku, was analyzed by the  X-ray satellite in six pointings,
 centered on the X-ray peak and offset by 
\timeform{14'} west, \timeform{30'}, \timeform{34'}, \timeform{44'}, and \timeform{60'}.
Because of its low background, Suzaku is the most sensitive X-ray satellite for observing 
 K$\alpha$ lines of Fe in the intracluster medium.
After obtaining accurate measurements of Fe lines, we studied the temperature
 structure of the intracluster medium, searched for possible bulk-motions, and 
measured the Fe abundance in the cluster.

The spectra of each extracted region were well fitted by the single-temperature APEC model,
and the two- or three- temperature APEC model did not improve $\chi^2$.
The temperatures derived from the observed ratios of K$\alpha$ lines of H-like and He-like Fe 
agree well with those from spectral fittings with the single-temperature APEC model.
Because this line ratio is a strong function of plasma temperature,
the observed consistency supports the accuracy of temperature measurements with Suzaku,
and constrains the temperature structure of ICM.

The observed values of the central energy of the He-like Fe line
of the center, \timeform{14'}, \timeform{30'} and \timeform{34'} offset regions
are constant within 500 ${\rm km~s}^{-1}$, which corresponds to the calibration error.
Because relative bulk velocities in the Coma cluster are smaller than 
 the sound velocity of the intracluster medium, 1500 km s $^{-1}$, 
we can verify the derived total mass in a cluster on the basis of the hydrostatic ICM equilibrium 
inside the \timeform{34'} offset region.
Significant bulk velocities were also not found in the \timeform{44'} offset region, 
which corresponds to the NGC 4839 subcluster.

The results on the temperature and velocity structure suggest that
the core of the Coma cluster is in a fairly relaxed state.  
This is consistent with models in which the
 non-thermal electrons relevant to the radio halo are accelerated by the intracluster 
turbulence rather than large-scale shocks. 
This is also consistent with the fact that numerical simulations show that
the turbulence motion is developed in the late phase of mergers.

The observed Fe abundance of the intracluster medium is almost constant at
 0.4 solar inside the \timeform{34'} offset region, and decreases with radius. 
The central abundance is slightly lower than that of other clusters or groups.
This indicates that central regions of the gas were mixed well 
during the past merger growth of the cluster.
\\

{\it Acknowledgements.} 
We wish to thank the referee for his/her suggestions on the paper.
The authors are grateful to all members of Suzaku for their
contributions in instrument preparation, spacecraft operation, software development, 
and in-orbit instrumental calibration. To analyze the data, 
we used the ISAS Analysis Servers provided by ISAS/C-SODA. 
N. O. acknowledges support by the
Ministry of Education, Culture, Sports, Science and
Technology of Japan, Grant-in-Aid for Scientific Research
No. 22740124.
CLS was supported in part by NASA Suzaku grants NNX08AZ99G, NNX09AH25G, and NNX09AH74G.

\end{document}